\documentclass[aps,amsmath,amssymb,pr,reprint,groupedaddress]{revtex4-1}

\usepackage{graphicx}
\usepackage{dcolumn}
\usepackage{bm}

\begin{document}

\title{
Universal scaling for recovery of Fourier's law 
\\
in low-dimensional solids under momentum conservation
}


\author{Dye SK Sato}
\email[]{
sato.daisuke.6r@kyoto-u.ac.jp
}
\affiliation{
Disaster Prevention Research Institute, Kyoto University, Gokasho, Uji, Kyoto 611-0011, Japan
}


\date{\today}

\begin{abstract}
Dynamic renormalization group (RG) of fluctuating viscoelastic equations is investigated to clarify the cause for numerically-reported disappearance of anomalous heat conduction (recovery of Fourier's law) in low-dimensional momentum-conserving systems. 
RG flow is obtained explicitly for simplified two model cases: a one-dimensional continuous medium under low pressure and incompressible viscoelastic medium of arbitrary dimensions.
Analyses of these clarify that the inviscid fixed point of contributing the anomalous heat conduction becomes unstable under the RG flow of non-zero elastic-wave speeds. 
The dynamic RG analysis further predicts a universal scaling of describing the crossover between the growth and saturation of observed heat conductivity, which is confirmed through the numerical experiments of Fermi-Pasta-Ulam $\beta$ (FPU-$\beta$) lattices.
\end{abstract}

\pacs{}

\maketitle

\section{Introduction}
Heat conduction has attracted wide attention with various motivations, such as testing irreversible thermodynamics~\cite{de2013non,seifert2012stochastic,casati1984one,takesue1990fourier}, 
exploring the innovative nano- and micro-scale materials~\cite{balandin2011thermal,gu2018colloquium},
and ultimately establishing a general theory of non-equilibrium statistical mechanics~\cite{forster1977large,faris1982large,lepri2003thermal,saito2007fluctuation}. 
In ordinary three-dimensional many-particle systems, the heat conduction is governed by a linear relation called {\it Fourier's law} between energy current ($J$) and the spatial gradient of temperature ($T$),  
\begin{eqnarray}
J=-\kappa \nabla T, \hspace{10pt}\kappa\propto N^0,
\end{eqnarray}
where $\kappa$, called heat conductivity, is an intensive variable independent of the number $N$ of particles.

Meanwhile, the long-range correlation of thermal fluctuations emerges due to the momentum-conservation in low-dimensional systems of dimension $d\leq2$,
and transport coefficients are there generally not intensive~\cite{kawasaki1965logarithmic,pomeau1975time,forster1977large,lepri2003thermal}. 
The heat conductivity can increase for such cases in proportion to the power of $N$ with a positive fractional exponent $\alpha$;
\begin{eqnarray}
\kappa \propto N^\alpha, \hspace {5pt} 0<\alpha <1.
\label{eq:heatconductionscaling}
\end{eqnarray}
This intriguing breakdown of Fourier's law, called {\it anomalous heat conduction}, has been investigated over recent decades~\cite{lepri1998anomalous,chang2008breakdown,lepri2016heat}.
This scaling suggests the absence of macroscopic descriptions of heat conduction in low-dimensional systems in the thermodynamic limit ($N\to\infty$).
The nano- and micro-scale experiments using carbon-nanotubes and graphene sheets have verified such scaling of the anomalous heat conduction as the natures of low-dimensional materials~\cite{chang2008breakdown,balandin2011thermal}.
With semi-macroscopic descriptions, called the fluctuating hydrodynamic equations~\cite{landau1986course}, theoretical studies have connected the anomalous heat conduction to
the Kardar-Parisi-Zhang (KPZ) universality class~\cite{narayan2002anomalous,lepri2003thermal,spohn2016fluctuating}.

However, recent numerical studies have been posing a considerable number of counterexamples to this anomalous heat conduction in the low-dimensional momentum-conserving systems~\cite{gendelman2000normal,jiang2016modulating,PhysRevE.94.012115,zhong2012normal,chen2012breakdown}. 
Ref.~\cite{gendelman2000normal} first indicated from a molecular-dynamic simulation that the heat conductivity can be intensive in a momentum-conserving one-dimensional model.
Some studies of the molecular dynamics~\cite{zhong2012normal,chen2012breakdown} confirmed Fourier's law in other one-dimensional momentum-conserving systems, 
although it was later criticized that the model size $N$ in their investigation was not large enough to study the asymptotic semi-macroscopic heat conduction~\cite{das2014heat}; 
and thus these following reports can be apparent, occurring due to the crossover between the ballistic transport $\alpha=1$ and the anomalous transport $\alpha=1/3$~\cite{chen2014nonintegrability}.
The asymptotic behavior is further investigated by quite recent studies~\cite{jiang2016modulating,PhysRevE.94.012115} 
with paradigmatic models of the anomalous heat conduction, such as FPU-$\beta$ lattices.
From the direct measurement of the size-dependent heat conductivity $\kappa(N)$ with different model sizes $N$, 
Ref.~\cite{PhysRevE.94.012115} reported that
the asymptotic anomalous growth Eq.~(\ref{eq:heatconductionscaling}) of the heat conductivity 
plateaus within larger sizes $N$ than a characteristic scale $N_*$,
\begin{eqnarray}
\kappa(N) \sim \kappa(N_*) [\min(N/N_*,1)]^\alpha,
\label{eq:predprevstud}
\end{eqnarray}
which is, so to speak, {\it recovery of Fourier's law}. 
In the case of FPU-$\beta$ lattices in Ref.~\cite{PhysRevE.94.012115}, 
the growth in $\kappa(N)$ plateaued around $N_*=10^6$; the scaling exponent $\alpha$ in Eq.~(\ref{eq:predprevstud}) is measured at an intermediate size scale $N<N_*$ and is met to the theoretically predicted value, $\alpha=1/3$ of the KPZ class~\cite{narayan2002anomalous,spohn2016fluctuating}. Ref.~\cite{PhysRevE.94.012115} also confirmed around $N= 10^2\sim 10^3$ that
the Green-Kubo formula~\cite{kubo2012statistical} can give system-size independent heat conductivity $\kappa_{GK}$ to the FPU-$\beta$ lattices under the periodic boundary conditions; 
\begin{eqnarray}
\frac{\partial}{\partial N}\kappa_{GK}=0.
\end{eqnarray}
Interestingly, despite their difference in sizes, measurements, and boundary conditions, $\kappa_{GK}$ meets the asymptotic value $\kappa(N_*)$ of the directly measured heat conductivity, i.e.,
\begin{eqnarray}
\kappa(N_*)\sim \kappa_{GK},
\label{eq:numericallyfoundquantificationofGKkappa}
\end{eqnarray}
under the same thermodynamic (pressure and temperature) conditions.
This coincidence, unexpected from the anomalous heat conduction yet satisfied in the case of the normal heat conduction~\cite{PhysRevE.94.012115,casati2003anomalous}, consistently indicates the recovery of Fourier's law at larger $N$ than the intermediate size of the anomalous heat conduction.

While these numerical reports systematically suggest the recovery of Fourier's law can add a surprising counterexample to the KPZ class,
the conflict remains between these molecular-dynamic studies and previously proposed semi-macroscopic theoretical studies. 
The recovery of Fourier's law concerns the thermodynamic limit, 
so the numerical approach is not enough to remove the suspicion~\cite{das2014role} of finite-size effects. 
Thermally activated dissociation is pointed out to be a possible origin of the recovery of Fourier's law in some model cases~\cite{gendelman2014normal,PhysRevE.94.012115} by inhibiting many-particle systems from obeying the continuum semi-macroscopic dynamics. 
However, that explanation is found to be not enough to describe all the recovery of Fourier's law, as its prediction of $N_*$ can be inconsistent with observations for some cases~\cite{PhysRevE.94.012115}.
Another study reinvestigated the dynamic RG flow of the fluctuating hydrodynamic equation~\cite{PhysRevE.94.012115}. It showed that a fixed point (called the inviscid fixed point) of contributing anomalous heat conduction becomes unstable when the pressure term is added to the noisy Burgers' equation of zero-pressure, and another fixed point (called the ballistic fixed point) emerges to characterize the recovery of Fourier's law as another universality class. 
This RG analysis may complement the previous theory of anomalous heat conduction, given that the previous studies are discussing/utilizing the existence of the (inviscid) fixed point of the anomalous heat conduction~\cite{narayan2002anomalous,mai2006universality} and are not studying its stability. 
In other words, as long as the cause for the recovery of Fourier's law is expressed as the instability of the inviscid fixed point, the recovery of Fourier's law can be compatible with the previous theories of the anomalous heat conduction. 
Consistently, as the RG analysis of Ref.~\cite{PhysRevE.94.012115} predicts, other recent molecular-dynamic studies reported the recovery of Fourier's law under non-zero pressure conditions, such as non-zero strain and anti-symmetric inter-particle potential~\cite{jiang2016modulating,barik2019temperature}.
Nevertheless, some previous semi-macroscopic theories already studied the stability of another anomalous transport (growing viscosity) against the pressure and strains by using the incompressible hydrodynamic equation~\cite{forster1977large}, 
so the conflict remains even between the suggestions of the RG analyses of 
Refs.~\cite{PhysRevE.94.012115,forster1977large}.

In this paper, we aim to provide a semi-macroscopic description for the recovery of Fourier's law in a manner compatible with the previous theoretical studies of the fluctuating hydrodynamic equations.
The analysis focuses on generalizing the previous RG analysis of Refs.~\cite{forster1977large,PhysRevE.94.012115} to the RG analysis of fluctuating viscoelastic (visco-elastodynamic) equations.
We will find that the existence of volumetric and solenoidal elastic-wave speeds is a key factor to relate previous theoretical RG analysis with the universal scaling for the recovery of Fourier's law.

This paper is organized as follows. The set of the fluctuating viscoelastic equations~\cite{mai2006universality} is introduced as an extension of the fluctuating hydrodynamic equations in {\it Setting} section. Dynamic RG analyses are executed in {\it Results} section. After the explicit development of the RG flow, the theoretical prediction of the RG flow is tested by numerical experiments with the FPU-$\beta$ lattices.
The relationship between presented results and previous ones is explored in {\it Discussion} section.

\section{Setting}
Formulation of the fluctuating hydrodynamic and viscoelastic equations is mentioned first. The models addressed in this paper are then introduced in an ordinary theoretical framework.

\subsection{Fluctuating Hydrodynamic Equations}
Suppose mass $\rho$, momentum $\rho{\bf v}$, and energy $e$ are conserved in microscopic scales, where ${\bf v}$ denotes the velocity.
Continuity equations there holds for their densities as
\begin{eqnarray}
\partial_t \rho +\partial_a (\rho v_a) &=&0
\nonumber \\
\partial _t (\rho v_a) + \partial_b (\rho v_a v_b) &=& \partial_b \sigma^\prime_{ab}
\label{eq:flfl}
\\
\partial_t e + \partial_a J^\prime_a &=& 0  
\nonumber 
\end{eqnarray}
where $v_a$ denotes the $a$ component of the velocity ${\bf v}$, 
and 
$\sigma_{ab}$ and $J^\prime_a$ respectively the $a,b$ component of the stress tensor 
and $a$ component of the energy current
in the fluctuating hydrodynamic equations;
$\partial_t$ and $\partial_a$ respectively represent partial derivatives in terms of time $t$ and the $a$ component of the location ${\bf x}$.

In the semi-macroscopic scales, 
the macroscopic equations are considered to dominate the collective average motions of many particles while being disturbed by non-negligible thermal fluctuations~\cite{landau1987fluid}.
The stress and energy currents are then
given as sums of deterministic parts and stochastic parts, 
\begin{eqnarray}
\begin{array}{lll}
\sigma^\prime_{ab}&=& -P\delta_{ab} +C^{vis}_{abmn}\partial_m v_n +s_{ab}
\\
J^\prime_a&=&ev_a-\sigma^\prime_{ab}v_b-\kappa \partial_ a T - g_a
\label{eq:flfl2}
\end{array}
\end{eqnarray}
where $P$ and $T$ respectively denote the pressure and temperature,
and $\kappa$ and $C^{vis}_{abmn}$ respectively the heat conductivity and the $a,b,m,n$ component of viscosity tensor $C^{vis}$; $s$ and $g$ denote the random stress and random heat current, respectively.  
The mass-density current include no dissipation, so no stochastic thermal effect arises there~\cite{landau1987fluid}.
Note that $P$ and $T$ are functions of mass density and internal energy density ($\rho, e -\rho v^2/2$).

Thermal fluctuations ($s, g$) are written by white noises as a consequence of the central limit theorem, 
and are governed by the fluctuation-dissipation relations (FDR)~\cite{kubo2012statistical};  
\begin{eqnarray}
\langle s_{ab}({\bf x},t)s_{mn}({\bf x^\prime},t^\prime) \rangle &=& 2 C^{vis}_{abmn} T \delta({\bf x}-{\bf x^\prime})\delta(t-t^\prime)
\nonumber \\
\langle g_a({\bf x},t)g_b({\bf x^\prime},t^\prime) \rangle &=& 2\kappa T^2 \delta_{ab}\delta ({\bf x}-{\bf x^\prime})\delta(t-t^\prime)
\label{eq:shearnoiseav}
\\
\langle s_{ab}({\bf x},t)g_c({\bf x^\prime},t^\prime) \rangle &=& 0 
\nonumber 
\end{eqnarray}
where $\langle \rangle$ denotes the noise average, and $\delta(t)$ and $\delta({\bf x})$ denote the Dirac delta functions of one- and $d$- dimensional spaces, respectively; $\delta_{ab}=1$ (when $a=b$) $=0$ (otherwise) is the Kronecker delta. 

Besides, the high wavenumber cutoff $\Lambda$ 
is assumed to express the non-continuum area of short-wavelength~\cite{landau1987fluid}.
The value of $\Lambda$ can change in the renormalization process~\cite{forster1977large}, as detailed later.

The following analysis follows the above theoretical framework of the semi-macroscopic motions, and the verification of it is out of the scope in this study. Please refer to Refs.~\cite{hoover2012computational,saito2011additivity} for the molecular-dynamic verification of this framework.

\subsection{Fluctuating Viscoelastic Equations}

Shear strain becomes an additional conserved order parameter in the viscoelastic materials that respond to shear deformation elastically~\cite{landau1986course,mai2006universality}. 
We here derive the fluctuating viscoelastic equations by focusing on a parameter range where such shear elastic response of media is relatively small.

The strain accumulates in an infinitesimal element of continuity
due to relative velocity difference from the surrounding other elements. It is written in the Lagrangian description as
\begin{eqnarray}
\frac{ D \epsilon_{ab}}{D t} = \frac 1 2 (\partial_a v_b +\partial_b v_a ),
\label{eq:evofstrain}
\end{eqnarray}
where $D/Dt:=\partial_t +v_a\partial_a$ is the Lagrangian differentiation operator and $\epsilon_{ab}$ denotes the $a,b$ component of the strain tensor.
Elastic order yields the shear stress components of the Hookean elastic stress $\sigma^{el}$ responding to the strain,
\begin{eqnarray}
\sigma^{el}_{ab}= C^{el}_{abmn}\epsilon_{mn},
\end{eqnarray}
where
$C^{el}_{abmn}$ is the $a,b,m,n$ component of stiffness tensor. 
The shear strain and stress are given by the traceless parts of the strain and stress tensors, respectively~\cite{de2013non}.

In the viscoelastic material, the elastic shear response perturbs the aforementioned stress $\sigma^\prime$ of obeying the viscous constitutive law. Such a perturbation can be additive, as long as the shear elastic response is small enough to be linearized.
Then adding the traceless part of the Hookean response to $\sigma^\prime$, we obtain the viscoelastic constitutive law of small elastic shear stress, 
\begin{eqnarray}
\sigma_{ab}=\sigma^\prime_{ab}+(\sigma^{el}_{ab}-\delta_{ab}\sigma^{el}_{cc}/d),
\label{eq:viscoelastoconstitutive}
\end{eqnarray}
where $\sigma_{ab}$ denotes the $a,b$ component of $\sigma$.
In terms of the volumetric part, Eq.~(\ref{eq:viscoelastoconstitutive}) subtracting the volumetric part of $\sigma^{el}$ is consistent with the thermodynamic definition of the pressure, that is the adiabatic response to the (reversible) volumetric change~\cite{landau1986course,landau1987fluid}.

After the replacement of $\sigma^\prime$ with $\sigma$, 
the fluctuating viscoelastic equations are obtained from the fluctuating hydrodynamic equations as
\begin{eqnarray}
\partial_t \rho +\partial_a (\rho v_a) &=&0
\nonumber \\
\partial _t (\rho v_a) + \partial_b (\rho v_a v_b) &=& \partial_b \sigma_{ab}
\nonumber \\
\partial_t e + \partial_a J_a &=& 0  
\nonumber\\
(\partial_t +v_c\partial_c ) \epsilon_{ab} &=&(\partial_a v_b+\partial_b v_a)/2
\label{eq:originalel}
\end{eqnarray}
with
\begin{eqnarray}
\sigma_{ab}&=& -P\delta_{ab} + (C_{abmn}-\delta_{ab}C_{ccmn}/d)\epsilon_{mn}
\nonumber \\
&&+C^{vis}_{abmn}\partial_m v_n+s_{ab}.
\label{eq:originalel2}
\\
J_a&:=&ev_a-\sigma_{ab}v_b-\kappa \partial_ a T - g_a
\nonumber 
\end{eqnarray}
Eqs.~(\ref{eq:originalel}) and (\ref{eq:originalel2}) are consistent with the one-dimensional viscoelastic equations derived from the thermodynamic discussions~\cite{mai2006universality}, after neglecting their vacancy diffusion terms.
In the following analysis,
we do not consider the vacancy diffusion despite that it is a possible cause of the recovery of Fourier's law in some cases~\cite{gendelman2014normal,PhysRevE.94.012115}. This is in order to focus on the effect of the elastic order in the fluctuating viscoelastic equations being our main concern.

It is noteworthy that there is no distinction between the viscoelastic and fluid bodies in one-dimensional systems where only the volumetric deformation exists.
The traceless elastic stress $\sigma^{el}_{ab}-\delta_{ab}\sigma^{el}_{cc}/d$ indeed becomes zero exactly at $d=1$, and elastic interactions are fully included in the pressure there.
The strict distinction is then unnecessary between the viscoelastic and fluid media, or more widely between the solid and liquid/gas, in one-dimensional systems in that sense. 
These hold for one-dimensional systems and quasi-one-dimensional systems such as carbon nanotubes becoming one-dimensional in a coarse-grained view.

Note that the change in volumetric strain $\epsilon_{aa}$ and the logarithmic change in the mass density $\rho$ are connected by the following relation:
\begin{eqnarray}
\frac{D}{Dt}[\log \rho + \epsilon_{aa}]=0.
\label{eq:volumetricchangeequivalence}
\end{eqnarray}
This can be obtained from Eqs.~(\ref{eq:flfl}) and (\ref{eq:evofstrain}).
Therefore, 
there can be indefiniteness in the formalism with respect to the volumetric changes. Nevertheless, 
newly imposed reversible stress $\sigma^{el}_{ab}-\delta_{ab}\sigma^{el}_{cc}/d$ in $\sigma$ does not contain $\epsilon_{aa}$ terms at least in the following analysis of isotropic homogeneous materials. 
Hence this study does not require any additional rules to avoid such indefiniteness.

\subsection{Two Simplified Models}
\label{sec:modelintroduction}
The anomalous transport is caused by the coupling of the thermal fluctuations incurred by the nonlinearity of the governing equations~\cite{forster1977large,pomeau1975time}.
Such growth of the transport coefficients is estimated from the analysis of the fluctuating motions of conserved quantities near an equilibrium state~\cite{forster1977large,kubo2012statistical,narayan2002anomalous} under the semi-macroscopic theoretical framework.
Two famous models, noisy Burgers fluids and stirred incompressible fluids, have been well studied near the equilibrium states in the context of anomalous transport~\cite{forster1977large}. 
I introduce modified models of these two representative models by adding the small perturbation of pressure and elastic shear stress, respectively, to study the lowest-order effect of the elasticity in the RG flow.

A static thermodynamic state of zero-momentum $(\rho,v,e,\epsilon)=(\rho_0,0,e_0,\epsilon_0)$ is considered as a reference state below, where $A_0:=\langle A \rangle$ represents the original reference value of a given variable $A$. The reference equilibrium state of zero-momentum is consistent with the starting point equation of the previous study~\cite{forster1977large}.

\subsubsection{Model Simplification}
Analyzed models follow widely-adopted approximations of 1) the constancy of the transport coefficients and 2) decoupling hypothesis~\cite{forster1977large,spohn2016fluctuating}. 
Regarding the constancy of the transport coefficients,
although the transport coefficients ($C^{vis},\kappa$) are thermodynamic quantities, 
previous studies have ordinarily neglected their variations caused by the fluctuations of the thermodynamic state (described by the thermodynamic state quantities such as temperature and pressure);
\begin{equation}
(C^{vis},\kappa)\simeq(C^{vis}_0,\kappa_0). 
\label{eq:adoptedappA}
\end{equation}
This approximation is verified by that the nonlinearity of the transport coefficients is irrelevant for analyzing the self-similar scaling (related to the fixed points of the RG flows) of the anomalous heat conduction~\cite{spohn2016fluctuating}.
Regarding the decoupling hypothesis,
the nonlinear contributions of the heat mode of linearized hydrodynamic equations are assumed to be irrelevant for the sound modes~\cite{spohn2016fluctuating}. The verification of the decoupling hypothesis is well summarized in Ref.~\cite{spohn2016fluctuating}. 
In the RG analysis near a zero-momentum reference state, as far as we study the linearized hydrodynamic equations with the streaming terms of the momentum- and energy-density currents, the diagram expansion for the renormalized transport coefficients under this decoupling hypothesis requires the energy-density current to behave just like a passive scalar~\cite{PhysRevE.94.012115}:
\begin{equation}
\partial_b( v_a\sigma_{ab})\to0, \left(\frac{\partial P}{\partial \delta e}\right)_{\delta \rho}\to 0.
\label{eq:adoptedappB}
\end{equation}
The current of the passive scalar has been investigated as a model system of the anomalous transport for decades~\cite{bedeaux1974renormalization,forster1977large}. 
Although Ref.~\cite{PhysRevE.94.012115} pointed this reduced form of the decoupling hypothesis for one-dimensional fluids (including one-dimensional Burgers' one), this holds for the incompressible and Burgers' fluids of arbitrary dimensions given that their diagram expansions are dimension-independent except prefactors~\cite{forster1977large}. The models investigated in this study have the same diagram expansions as these fluidic models, so Eq.~(\ref{eq:adoptedappB}) is the reduced form of the decoupling hypothesis there as well. 

Besides, in the presented RG analysis, the cubic- or higher-order order fluctuations are dropped as in the previous study~\cite{forster1977large}. 
This is a working hypothesis to obtain the differential RG flow (even in the previous analysis~\cite{forster1977large}), so this is mentioned later at the beginnings of the RG analyses.
Note that this hypothesis has been examined theoretically~\cite{eyink1994renormalization}, and is widely accepted in the theoretical studies of the anomalous transport as it is adopted even in the mode-coupling analysis~\cite{spohn2016fluctuating}.

Homogeneity and isotropy of the medium are adopted widely in the anomalous-transport studies~\cite{forster1977large}. 
The viscosity tensor is there parametrized as
\begin{eqnarray}
C^{vis}_{abmn}=
\eta\delta_{ab}\delta_{mn}+
\zeta(\delta_{am}\delta_{bn}+\delta_{an}\delta_{bm})
 \label{eq:isohomosto}
\end{eqnarray}
with
shear viscosity $\zeta$ and volume viscosity $\eta+2d^{-1}\zeta$. 
The stiffness tensor becomes
\begin{eqnarray}
C^{el}_{abmn}=\tilde \lambda \delta_{ab}\delta_{mn}+
\mu(\delta_{am}\delta_{bn}+\delta_{an}\delta_{bm}),
\end{eqnarray}
where $\tilde \lambda$ and $\mu$ respectively denote Lame's first and second parameters, and $\mu$ corresponds to the rigidity of the medium.
Note that the assumption of the isotropy is unnecessary for studying one-dimensional cases~\cite{narayan2002anomalous,spohn2016fluctuating} where any tensors reduce to a scalar.

To get results comparable with those previous studies, I here focus on the minimum changes of these fluidic models occurring due to the elasticity of the media. 
First, I drop the nonlinear terms multiplied with the stiffness tensor, e.g.,
\begin{equation}
C^{el}\simeq C^{el}_0.
\end{equation}
Second, I adopt the zero shear strain or negligibly small shear strain for the reference equilibrium state,
\begin{equation}
\epsilon_{0,ab}-\delta_{ab}\epsilon_{0,cc}/d\to0.
\end{equation}
Near the zero-shear-strain states, the dependence of the temperature and pressure on traceless strain starts from the second order of fluctuations with the stiffness tensor. 
This is because traceless-strain dependencies of them come from the internal strain energy (contributed from the square of the strain) or the second and third invariants of stress (on the square order of the strain), due to the requirement of the coordinate-rotation invariance of the governing equations.
Related second-order fluctuations multiplied with the stiffness tensor are dropped in the following analysis as the nonlinearity related to the elastic shear stress.
These linearizations in terms of shear elastic motions are to investigate the lowest-order effect associated with the shear elastic response.

As above, the assumptions adopted in the model equations of this study are
the assumptions inherited from the previous theoretical studies and the linearization of the elastic effects. Given that the assumptions of previous studies are well investigated by the previous studies themselves over decades (as summarized in Ref.~\cite{spohn2016fluctuating}), 
the crucial assumption of this study will be the linearization concerning the elastic response that drops the nonlinear terms proportional to the stiffness tensor. 
The aim of this paper is to study how the RG flow of the previous studies is perturbed by the elastic effect they did not consider. 
The linearized elastic effect represents the most dominant part of such perturbation. I focus on it, and the limitation of this study due to the neglected higher-order elastic perturbation will be examined in the discussion section.

For later analytical simplicity, 
I introduce the displacement $\psi$ with setting its reference value at 0 without loss of generality, and then Hookean response is expressed as
\begin{equation}
C^{el}_{abmn}\epsilon_{mn}=C^{el}_{abmn}\partial_m\psi_n.
\end{equation}

\subsubsection{Incompressible Viscoelastic Materials}
\label{sec:introducingmodel1}
After these simplifications, one simple model is obtained with the approximation of incompressibility,
\begin{equation}
\delta\rho=0,
\end{equation}
where $\delta$ represents the index of the fluctuations giving $\delta A:=A-\langle A\rangle $ to a variable $A$. 
This condition is equivalent with
\begin{equation}
\partial_a\psi_a=\epsilon_{aa}=0,
\end{equation}
given Eq.~(\ref{eq:volumetricchangeequivalence}).
Under such a condition of incompressibility, 
with aforementioned model simplification and near the aforementioned equilibrium states, 
 original equations (\ref{eq:originalel}) and (\ref{eq:originalel2}) reduce to 
\begin{eqnarray}
&\partial_t \psi_a = v_a ,\,\, \partial_a v_a=0
\nonumber\\
&\partial _t v_a + \partial_b(v_a v_b) = -\partial_a \delta p + Y_0 \Delta \psi_a + \nu_0\Delta v_a +\partial_b s^\prime_{0ab}
\nonumber
\\
&\partial_t \delta e +\partial_a(\delta e  v_a) = D \Delta  \delta e +  \partial_a g_{0a}
\label{eq:incompelpassivescalar}
\end{eqnarray}
up to the second-order fluctuations, where $D:=\kappa(\partial T/\partial e)_\rho$ and $\nu_0:=\zeta_0/\rho_0$ denote thermal diffusivity and kinetic viscosity, respectively; variables $p:=P/\rho, Y:=\mu/\rho,s^\prime :=s/\rho_0$ are introduced for brevity.
$Y$ represents the square of the transverse wave speed.
The functional form of $\delta p$ is given by a rewritten incompressible condition: 
\begin{equation}
\partial_av_a=0.
\label{eq:incompressiblecondwithvdiff}
\end{equation}
The difference of this model equation from that for the incompressible fluids in Ref.~\cite{forster1977large} fully comes from the elastic shear response $Y_0\Delta \psi_a$ and the energy current fluctuations $g_{0a}$. 
Whereas $g_{0a}$ is neglected in the previous study~\cite{forster1977large} for technical simplicity, 
the following results of the RG flows are independent from the presence or absence of $g_{0a}$. The difference in $g_a$ is thus irrelevant to compare the following analysis and that in Ref.~\cite{forster1977large}. The intrinsic difference between the previous and following RG analyses is the elastic shear contribution $Y_0\Delta \psi_a$ only.

This fluctuating incompressible viscoelastic equation represents the minimum change of the fluctuating incompressible fluids induced by the elastic shear response. 
The following analysis is based on this fluctuating incompressible viscoelastic equation near $d=2$, 
where the heat conductivity is shown to diverge for the case of the stirred incompressible fluids (corresponding to $Y_0=0$)~\cite{forster1977large}. At $d=1$, incompressible approximation only gives a linear heat diffusion equation with a trivial rigid-body solution ${\bf v}={\bf 0}$ and fails to capture the anomalous heat conduction even in the previous study~\cite{forster1977large}. So we consider another model next below. 

\subsubsection{One-Dimensional Materials Under Low Pressure}
\label{sec:introducingmodel2}
To study one-dimensional cases, 
a useful set of model equations is the tandem of the diffusion equation of 
a passive scalar and the noisy Burgers' equation~\cite{forster1977large,drossel2002passive,PhysRevE.94.012115}.

By adding energy-density-independent small pressure perturbation to the noisy Burgers' equation, 
a minimal model is obtained to study the effect of pressure in the RG flow of one-dimensional fluctuating hydrodynamic equation;
\begin{eqnarray}
\partial_t \delta \rho +\partial_x u&=&0
\nonumber
\\
\partial_t  u + \rho_0^{-1}\partial_x u^2 &=& -Y_0\partial_x\delta \rho+\nu_0\Delta u+\partial_x s
\label{eq:1dburgwp}
\\
\partial_t \delta e+ \rho_0^{-1}\partial_x u\delta e&=& D\Delta \delta e +\partial_x g.
\nonumber
\end{eqnarray}
where $u:=\rho v$ denotes the momentum.
This set of equations is obtained from a one-dimensional hydrodynamic equations as in a similar manner to the incompressible cases. Please refer to Ref.~\cite{PhysRevE.94.012115} for details.
The variable $\nu:=(\eta_0+2d^{-1}\zeta_0)/\rho_0$ represents the volumetric kinetic viscosity as in the Burgers fluids and is not the kinetic shear viscosity appearing in the incompressible fluids. Then $\nu$ consistently represents the characteristic viscosity in both of the presented model equations.
Similarly, the definition of $Y:=(\partial P)/(\partial \rho)_e$ is here the square of the longitudinal wave speed (sound speed), and is different from the case of the incompressible fluids where $Y$ represents the square of the transverse wave speed. In both models, $Y$ represents the square of the characteristic wave speed. 
As in the incompressible cases, the difference from the previous study of the noisy Burgers equation (associated with $Y_0=0$) is only $Y_0$ for the following RG analysis.

\section{Results}
The RG flows of the semi-macroscopic model equations are presented in this section.
The prediction of RG flows is tested subsequently by the microscopic molecular-dynamic simulations of the FPU-$\beta$ lattices.

\subsection{Dynamic Renormalization-Group Analysis}

The RG flow of the fluctuating incompressible viscoelastic equations is studied particularly near $d=2$.
At $d=1$, that of Eq.~(\ref{eq:1dburgwp}) is investigated. 

\subsubsection{Dynamic Renormalization-Group Analysis of Incompressible Viscoelastic Materials}
We begin the calculation of the dynamic RG flow with erasing the pressure and displacement from the governing equations in a non-perturbative manner.
The following analysis is executed in the Fourier domain with the Fourier transform $f({\bf k},\omega)=\int d{\bf x}dt\exp[i({\bf k\cdot x}+\omega t)]f({\bf x},t)$, where ${\bf k}$ and $\omega$ denote the wavenumber vector and angular frequency, respectively. 
The strain is expressed by using the velocity, and the functional form of the pressure is determined from an incompressible condition ($\partial_a v_a=0$). Consequently, the original equations reduce to
\begin{eqnarray}
v_a &=& G \hat f_a - i\lambda G P_{abc} (v_b*v_c)
\\
\delta e &=& gf^\prime - i \lambda gk_a[v_a*(\delta e+\rho_0\delta p)]
\end{eqnarray}
in the Fourier space with
\begin{eqnarray}
G&:=&\left[i\omega + (\nu_0-i Y_0/\omega) k^2\right]^{-1}
\nonumber\\
g&:=&(i\omega +D_0 k^2)^{-1}
\label{eq:greenfuncandpre}
\\
\delta p&=&-k^{-2}[ik_cf_c+k_bk_c\lambda(v_b*v_c)], 
\nonumber
\end{eqnarray}
where $f_a:=\partial_b s_{0ab}/\rho_0,\hat f_a:=P_{ab}f_b, f^\prime:=\partial_a g_a$ express noises, 
$P_{abc}:=(P_{ab}k_c+P_{ac}k_b)/2$ is defined according to the convention~\cite{forster1977large} with $P_{ab}({\bf k}):=\delta_{ab}-k_ak_b/k^2 $, 
and
$\lambda$ is a nonlinear intensity factor as in previous studies~\cite{forster1977large}, which expresses the smallness of the fluctuations; $(a*b)({\bf k},\omega):=\int d{\bf q} d\Omega a({\bf k}-{\bf q},\omega -\Omega)b({\bf q},\Omega)$ expresses the convolution of the arbitrary functions $a$ and $b$ in the Fourier space.

Among the three noise terms, $\hat f_a$ only contributes to the renormalizations in the following analysis 
and
has the following properties for isotropic viscous tensor Eq.~(\ref{eq:isohomosto}),
\begin{eqnarray}
\langle \hat f_a({\bf k},t)\hat f_b({\bf k}^\prime,t^\prime) \rangle &=& 
\frac{\Sigma_0k^2P_{ab}({\bf k})}{(2\pi)^{d+1}}\delta ({\bf k}+{\bf k}^\prime)\delta(t+t^\prime),
\nonumber
\\\Sigma&:=&2\rho^{-1}\nu T.
\label{eq:noisefhat}
\end{eqnarray}
$P_{ab}$ projects $f_b$ to the incompressible solenoidal motions of $\hat f_a$. 

We next calculate the RG flow in a differential manner.
The following two procedures repeat alternately in the dynamic renormalization group. 

One is eliminating short-wavelength fluctuations within the spherical shell $\Lambda e^{-l}<k<\Lambda$ in the wavenumber domain. 
Their contributions to the longer-wavelength motions can be calculated from averages taken over short-wavelength fluctuations in the shell. 
Such contributions are renormalized to the phenomenological constants in the above set of the governing equations as 
\begin{equation}
(\nu_0,\Sigma_0,D_0)\to (\nu_R,\Sigma_R,D_R), 
\end{equation}
where the subscript $R$ is an index to represent a renormalized value. Shifts from the original values to the renormalized values are on the order of the thickness of the shell, that is, of $\mathcal O(\Lambda l)$. The lowest order $\mathcal O(\Lambda l)$ contribution to the renormalized phenomenological constants is calculated from one-loop solutions~\cite{forster1977large} as $n$-loop contribution is of $\mathcal O((\Lambda l)^n)$. 

For obtaining the RG flow, the cubic or higher orders (such as $v^3$) in the renormalized governing equation are eliminated for the case of the incompressible hydrodynamic fluctuations~\cite{forster1977large}. Its validity is investigated in Ref.~\cite{eyink1994renormalization} in a nonperturbative manner with the help of the dimensional analysis. 
Ref.~\cite{eyink1994renormalization} showed that these higher orders become marginal at $d<2$, and irrelevant at $d\geq 2$, for the stirred incompressible fluids (called Model A of Ref.~\cite{forster1977large} in Ref.~\cite{eyink1994renormalization}). 
The physical applicability of the incompressible fluid model is strictly for $d\geq 2$ giving well-defined solenoidal fields, so the RG flow of the stirred incompressible fluids give the exact results even with dropping the higher orders, as far as within its applicable dimensions~\cite{eyink1994renormalization}. 
The model investigated in this subsubsection has the same diagrams as the stirred incompressible fluids, except the difference in the functional form of the Green's function $G$ for the momentum fields, so the dimensional analysis provided by Ref.~\cite{eyink1994renormalization} holds for this model as well. We thus drop the third- and higher-order terms in the incompressible viscoelastic equations below as in Ref.~\cite{forster1977large}.

The other procedure is rescaling. The wavenumber is there rescaled as 
\begin{equation}
k \to k^\prime := ke^l
\end{equation}
so to keep the total spherical diameter at $\Lambda$ in the wavenumber domain. The angular frequency is also scaled as
\begin{equation}
\omega \to \omega^\prime:=\omega e^{\int dl z(l)}.
\label{eq:rescalingoffreq}
\end{equation}
For obtaining the scaling for the leading-order term of the governing equations, $z$ is determined so as to keep the divergent coefficient finite. In the dynamic RG flow of the fluctuating hydrodynamic equation~\cite{forster1977large}, $z$ is chosen to rescale $\nu_R$ to $\nu_0$ so as to keep the substantial kinetic viscosity constant. I follow this definition of $z$ tentatively, and the validity of this choice is considered later. Last, the variables $(v,\delta e)$ are also rescaled to keep the characteristic intensity of the fluctuations (temperature) constant as $(v,\delta e)\to(v^\prime,\delta e^\prime):=(v,\delta e)\exp[-\int dl(z+d/2)]$.
These rescalings modify the values of renormalized coefficients in the renormalized and rescaled governing equations as 
\begin{equation}
(\nu_R,\Sigma_R,D_R)\to (\nu_R,\Sigma_R,D_R)e^{\int dl(z-2)}.
\label{eq:rescalingofcoeffs}
\end{equation}

Their alternate repetitions regarding respective shells of the infinitesimal thicknesses ($\Lambda l\to\Lambda dl$) determine the flow of the dynamic renormalization-group of the phenomenological coefficients 
$(\nu,\Sigma,D)$ in the renormalized and rescaled equations. The flow is given by 
the differential form of the renormalization group:
\begin{eqnarray}
&&\frac{d \nu}{dl}= \nu(z-2+A_d\bar \lambda^2)
\nonumber
\\
&&\frac{d \Sigma}{dl}= \Sigma(z-2+A_d\bar \lambda^2)
\nonumber
\\
&&\frac{d \lambda}{dl}= \lambda(z-1-d/2)
\label{eq:flowofincompelpassivescalarforarbitraryz}
\\
&&\frac{d Y}{dl} = Y(2z-2)
\nonumber
\\
&&\frac{d D}{dl} = D\left(z-2+\frac{(d-1)\tilde K_d}{\bar\kappa(1+\bar \kappa)+\bar Y}\bar\lambda^2\right)
\nonumber
\end{eqnarray}
where
$\bar\lambda:=\lambda \sqrt{\Sigma\Lambda^{d-2}/\nu^3},\bar Y:=Y/(\nu\Lambda)^2, \bar \kappa:=\kappa(\partial P/\partial e)_\rho/\nu $ are non-dimensionalized coefficients introduced for brevity, and $l$ in this set of differential equations parametrizes the accumulations of eliminated (and rescaled) shells of the infinitesimal thickness; 
two dimension-dependent non-dimensional constants $A_d$ and $K_d$ are introduced as in Ref.~\cite{forster1977large} as
\begin{eqnarray}
A_d&:=&\frac{d^2-2}{(d^2+2d)(2\sqrt \pi)^d\Gamma(d/2)}
\\
\tilde K_d&:=&[2(2\sqrt\pi)^{d}\Gamma(d/2+1)]^{-1},
\label{eq:valueofKd}
\end{eqnarray}
where $\Gamma(\cdot)$ denotes the Gamma function. Note that $\tilde K_d$ corresponds to $K_d/d$ in Ref.~\cite{forster1977large}; 
although the specific form of $\tilde K_d$ ($K_d/d$) is different by factor 2 between Eq.~(\ref{eq:valueofKd}) and that of Ref.~\cite{forster1977large}, it is not relevant for the following discussions, and so I do not discuss it here.
The parts proportional to $\lambda^2$ ($\bar\lambda^2$) in Eq.~(\ref{eq:flowofincompelpassivescalarforarbitraryz})
correspond to the renormalization contributed from the short-wavelength fluctuations, and the others are associated with the rescaling.

The value of $z$ is chosen to keep $\nu$ invariant as $z=2-A_d\bar\lambda^2$, and the flow reduces to
\begin{eqnarray}
&&\frac{d \bar \lambda}{dl}= \bar \lambda(1-d/2-A_d\bar\lambda^2)
\nonumber
\\
&&\frac{d \bar Y}{dl} = \bar Y(2-2A_d\bar \lambda^2)
\label{eq:RGincompel}
\\
&&\frac{d \bar \kappa}{dl} = 
\bar \kappa\bar \lambda ^2\left( -A_d +
\frac{(d-1)\tilde K_d}{\bar\kappa(1+\bar \kappa)+\bar Y}\right).
\nonumber
\end{eqnarray}
This is the RG flow for the incompressible viscoelastic equations Eqs.~(\ref{eq:incompelpassivescalar}) and (\ref{eq:incompressiblecondwithvdiff}) of keeping the kinetic viscosity constant. 
The value of $\bar\kappa$ $[=\kappa(\partial P/\partial e)_\rho/\nu]$ expresses the size-dependence of the ratio between $\kappa$ and $\nu$, because the heat capacity 
$(\partial T/\partial e)_\rho$ (of constant mass density) is an intensive thermodynamic quantity and thus size-independent.

In $d<\sqrt 2$, this flow does not have any nontrivial fixed points with a real $\bar \lambda$ value, and thus cannot predict anomalous transports seen in the simulations at $d=1$. 
It would reflect the rigid body behavior ${\bf v}={\bf 0}$ in one-dimensional incompressible systems. In the case of $d=1$, nonlinear couplings ($v_av_b=0,\delta ev_a=0$) cancels in the governing equations, and the energy density obeys the normal diffusion equation, as noticed from Eqs.~(\ref{eq:incompelpassivescalar}) and (\ref{eq:incompressiblecondwithvdiff}). 
This is the same problem as for the stirred incompressible fluids and should be regarded as the limitation of the incompressible models.
The following analysis of the RG flow Eq.~(\ref{eq:RGincompel}) focuses on $\sqrt 2<d\leq 2$, where the fluctuating incompressible hydrodynamic equations predict the anomalous heat conduction. 

In $d<2$, the trivial fixed point $\bar \lambda=0$ of  Eq.~(\ref{eq:RGincompel}) is unstable when the flow is perturbed to the $\bar \lambda$ direction. Therefore,
the value of $\bar \lambda$ is noticed to converge to $\bar \lambda_*:=\sqrt{(1-d/2)/A_d}$, being a unique linearly stable fixed point for the differential RG flow of $\bar\lambda$, at $l\gg1$, given that the renormalization starts with $\lambda>0$. 
This convergence exponentially progresses as $l$ accumulates. 
At $\lambda =\lambda_*$, the flow has two fixed points. One is the fixed point for zero rigidity limit $\bar Y\to0$,
\begin{eqnarray}
(\mu/(\zeta \Lambda)^2, \kappa c_T/\nu )=(0,C_{Id}(>0)),
\label{eq:inviscidFPforincomp}
\end{eqnarray}
where the heat capacity $(\partial T/\partial e)_\rho$ of the constant mass density is rewritten as $c_T$ and non-dimensional constants $\bar Y$ and $\bar \kappa$ are rewritten as $\mu/(\zeta \Lambda)^2$ and $\kappa c_T/\nu$ with dimensional constants for an explanatory purpose;
the positive constant $C_{Id}$ is estimated as 
\begin{equation}
C_{Id}\sim\sqrt{(d-1)\tilde K_d/A_d}
\end{equation} by using $(d-1)\tilde K_d/A_d\gg1$. 
This fixed point is the previously known fixed point of the anomalous heat conduction~\cite{forster1977large}. 
The flow reaches to this fixed point Eq.~(\ref{eq:inviscidFPforincomp}) if $\bar Y_0=0$, but it is generally unstable against the perturbation of the imposed rigidity. For the case of the non-zero rigidity, $\bar Y$ grows exponentially,
\begin{equation}
\bar Y \sim \bar Y_0e^{ld},
\end{equation}
at $l\gg1$. Note $2-2A_d\bar \lambda^2\sim d$. By using this exponential growth of $\bar Y$, the following exponential decay of $\bar \kappa$ is obtained as a solution for $\bar \kappa$ at $l\gg1$ in the case of $\bar Y_0>0$: 
\begin{equation}
\bar \kappa\sim \exp(-A_d\bar \lambda^2l).
\end{equation}
The linearly stable fixed point for $\bar Y_0\geq 0$ is then found to be 
\begin{eqnarray}
(\mu/(\zeta \Lambda)^2, \kappa c_T/\nu )=(\infty ,0).
\end{eqnarray}
The relation $\kappa c_T/\nu=0$ means the breakdown of the hyperscaling between the kinetic viscosity and the heat conductivity, and is clearly corresponding to the recovery of Fourier's law.

The case of $d=2$ is delicate because the nontrivial fixed point of $\bar \lambda$ degenerates to the trivial one in Eq.~(\ref{eq:RGincompel}) as $\bar\lambda_*\to0$. For $d=2$, the $l$-dependence of $\bar \lambda$ is there obtained as
\begin{eqnarray}
\bar \lambda=1/\sqrt{2A_d l+1},
\label{eq:lambdabardecay}
\end{eqnarray}
with using $\lambda_0=1$. 
Its converging rate to the fixed point is much slower than the exponential speed in $d<2$.
This slow inverse-square-root decay yields the necessity to consider the transient behavior of $\bar \lambda$ in the following analysis for $\bar\kappa$.
The previous work~\cite{forster1977large} revealed a nontrivial fixed point in $d=2$ with $\bar Y=0$,
\begin{eqnarray}
(\mu/(\zeta \Lambda)^2, \kappa c_T/\nu )=(0,C_{I2}).
\end{eqnarray}
In our analysis, $C_{I2}$ is given as
\begin{eqnarray}
C_{I2}:=(1+\sqrt{17})/2.
\end{eqnarray}
Note that because the coefficient $\tilde K_d$ of ours is slightly different (by factor 2) from that of Ref.~\cite{forster1977large}, the value of $C_{I2}$ in the above estimate deviates from that of Ref.~\cite{forster1977large}, yet discussing such a subtle factor is out of the scope of this study as mentioned earlier. 
Meanwhile, the exponential growth of $\bar Y$ occurs even at $d=2$ for the initial condition $\bar Y_0\neq 0$; 
\begin{equation}
\bar Y\sim \bar Y_0e^{2l}.
\label{eq:growingYbarin2dincomp}
\end{equation}
In this case of $\bar Y_0\neq 0$, 
the evolution of $\bar\kappa$ 
transitions around a particular value $l_*$ of $l$ ($l\sim l_*$) such that
\begin{equation*}
\bar \kappa(\bar \kappa+1)\sim \sqrt {\bar Y}.
\end{equation*} 
The terms in both hands are in the denominator of the third term in the flow for $\bar \kappa$ in Eq.~(\ref{eq:RGincompel}).
At $l\lesssim l_*$, $\bar Y$ is negligible there, 
and $\bar \kappa$ converges to $C_{I2}$ as in the case of $\bar Y=0$;
\begin{equation}
\bar \kappa \sim C_{I2}.
\label{eq:RGincompinviscid}
\end{equation}
This convergence is non-asymptotic, and so we cannot find simple expression to describe $\bar \kappa$ approaching $C_{I2}$. 
In contrast to $l\lesssim l_*$, the third term in Eq.~(\ref{eq:RGincompel}) vanishes when $l$ is large enough to give $l\gg l_*$, i.e., ${\bar Y}\gg\bar \kappa(\bar \kappa+1)\sim C_{I2}(C_{I2}+1)$. The corresponding asymptote of $\bar \kappa$ at $l\gg l_*$ is estimated as
\begin{eqnarray}
\bar \kappa \sim \bar \kappa_*/\sqrt{l/l_*}
\label{eq:RGincompofhc_nonreg}
\end{eqnarray}
with a constant $\bar \kappa_*$, by using $\bar \lambda \sim 1/\sqrt{2A_d l}$ obtained from Eq.~(\ref{eq:lambdabardecay}) at $l \gg 1/(2A_d)$.  
The value of $\bar \kappa_*$ is expected to be near $C_{I2}$ given the converging nature of $\bar \kappa$ to $C_{I2}$ at $l\lesssim l_*$.
Eqs.~(\ref{eq:growingYbarin2dincomp}) and (\ref{eq:RGincompofhc_nonreg}) indicate that the flow converges to the fixed point
\begin{eqnarray}
(\mu/(\zeta \Lambda)^2, \kappa c_T/\nu )=(\infty,0)
\label{eq:maybeballisticfixedpoint2D}
\end{eqnarray}
at $l\gg l_*$.
The recovery of Fourier's law thus remains even in $d=2$ as long as $\bar Y>0$, meaning the non-zero elastic shear resistance (non-zero rigidity). 

The following is convenient for expressing two asymptotes of 
Eq.~(\ref{eq:RGincompinviscid}) for $l\lesssim l_*$ and 
Eq.~(\ref{eq:RGincompofhc_nonreg}) for $l\gg l_*$: 
\begin{eqnarray}
\bar \kappa \sim \bar \kappa_*/(1+\sqrt{l/l_*}).
\label{eq:RGincompofhc}
\end{eqnarray}
The factor $1$ in the denominator is to regularize this expression around $l\lesssim l_*$, and Eq.~(\ref{eq:RGincompofhc}) reduces to Eq.~(\ref{eq:RGincompinviscid}) at $l\lesssim l_*$ and 
Eq.~(\ref{eq:RGincompinviscid}) at $l\gg l_*$. 

\begin{figure*}[tbp]
   \includegraphics[width=80mm]{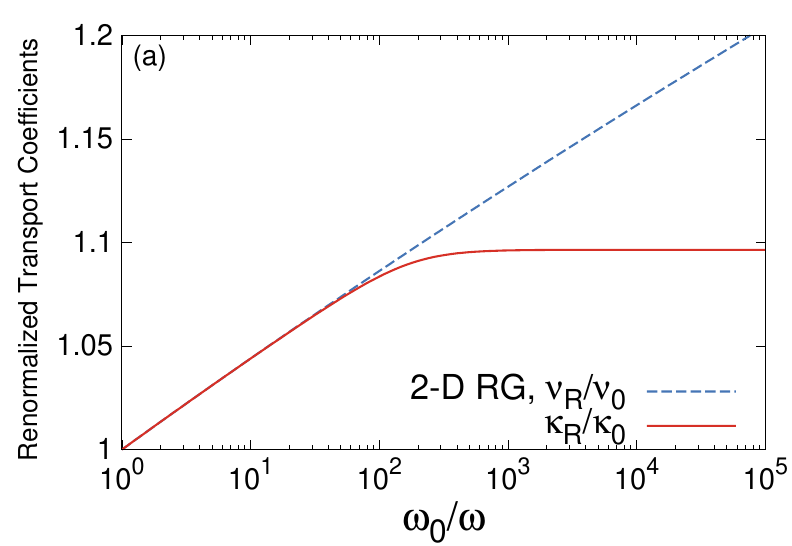}
   \includegraphics[width=80mm]{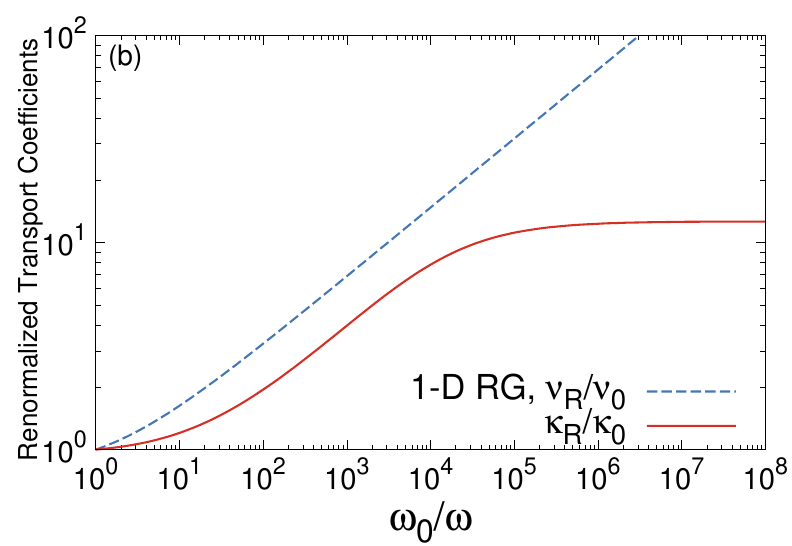}
  \caption{
Predicted ratios  
($\nu_R/\nu_0$ and $\kappa_R/\kappa_0$) of
the renormalized transport coefficients ($\nu_R$ and $\kappa_R$) to the bare transport coefficients ($\nu_0$ and $\kappa_0$)
in the angular frequency scale $\omega$, indicating the breakdown of the hyperscaling between the kinetic viscosity and the heat conductivity; 
$\omega/\omega_0=\exp(-\int dl z)$, $\nu_R/\nu_0=\exp[-\int dl(z-2)]$, 
$\kappa_R/\kappa_0=(\bar \kappa/\bar \kappa_0)\nu_R/\nu_0$ are given by the flows of the dynamic renormalization group as the functions of $l$ and the selected initial conditions $(\bar\lambda_0,\bar\kappa_0,\bar Y_0)$.
(a) 
Frequency-dependence of the renormalized transport coefficients in two-dimensional incompressible viscoelastic materials, predicted by the RG flow Eq.~(\ref{eq:RGincompel}) of $d=2$, for the case of an initial condition $(\bar\lambda_0,\bar \kappa_0,\bar Y_0)=(1,1,10^{-4})$ .
(b)
Frequency-dependence of the renormalized transport coefficients in one-dimensional materials under low pressure, predicted by the RG flow Eq.~(\ref{eq:RGburgwp}), for the case of an initial condition $(\bar\lambda_0,\bar \kappa_0,\bar Y_0)=(1,1,10^{-3})$. 
  }
  \label{fig:one}
\end{figure*}

The above analysis of the RG flow for the rescaled coordinates can describe the growth of the transport coefficients in the original scale~\cite{forster1977large}. 
Since it requires a transformation rule between the original and rescaled coordinates, I first describe that rule by following the formalism developed by the previous studies~\cite{forster1977large,narayan2002anomalous}. After deriving the related simple asymptotes analytically, 
we see the exact scaling of transport coefficients in a numerical way. [Fig.~\ref{fig:one} (a)]. I here focus on $d=2$ of interest.

Variables in the original coordinate are obtained as products of the scaled variables and scaling factors. For example, a given wavenumber $k(l)$ at finite $l$ is scaled as $k(l)=k(0) \exp(-l)$ from the original wavenumber k(0). The cutoff wavenumber is also as $\Lambda(l)=\Lambda_0\exp(-l)$. The angular frequency is scaled as $\omega(l)=\omega(0) \exp[-\int dl z(l)]$. 

The kinetic viscosity $\nu(l)$ at given $l$  in the RG flow is the transport coefficient renormalizing the eliminated fluctuations that have belonged to the shorter wavelength $k>\Lambda (=\Lambda_0)$ in the rescaled coordinate. 
Therefore, with rescaling, $\nu(l)$ becomes the renormalized kinetic viscosity at the wavenumber $\Lambda(l)$ in the original scale. 
The renormalized kinetic viscosity $\nu_R(\Lambda(l))$ of the original coordinate is 
then given from $\nu(l)$ with its scaling factor;
\begin{equation}
\nu_R(\Lambda e^{-l})=\nu(l)e^{-\int dl [z(l)-2]}.
\end{equation}
Note $\Lambda(l)=\Lambda e^{-l}$.
Further considering that $\nu(l)$ is unchanged from $\nu_0$ in the investigated RG flow [Eq.~(\ref{eq:RGincompel})], 
we obtain
\begin{equation}
\nu_R(\Lambda e^{-l})=\nu_0e^{-\int dl [z(l)-2]}.
\label{eq:nuscomputationalrule}
\end{equation}
Besides,
$\kappa_R c_T$ is given as the product of $\bar \kappa$ and $\nu_R$. By considering the constancy of $c_T$ mentioned earlier, and that the rescaling factors of $\nu$ and $D(=\kappa c_T)$ are the same in the RG flow [Eq.~(\ref{eq:rescalingofcoeffs})], we obtain
\begin{equation}
\kappa_R(\Lambda e^{-l})c_T
=
\bar \kappa(l)
\nu_R(\Lambda e^{-l}).
\end{equation}
It gives
\begin{equation}
\kappa_R(\Lambda e^{-l})/\kappa_0
=
[\bar \kappa(l)/\bar\kappa_0]
\nu_R(\Lambda e^{-l})/\nu_0.
\label{eq:kappascomputationalrule}
\end{equation}

Let us obtain the asymptotes for $\nu_R$ and $\kappa_R$.
Because $z=2-A_d\bar \lambda^2$ asymptotically reaches to $2-1/(2l)$ due to 
$\bar \lambda \sim 1/\sqrt{2A_dl}$ [Eq.~(\ref{eq:lambdabardecay}) at $l\gg1$], 
the renormalized kinetic viscosity at $l\gg1$ is given as 
\begin{eqnarray}
\nu_R(\Lambda e^{-l})\sim\nu_0 \sqrt{l},
\label{eq:2DexponentofviscosityA}
\end{eqnarray}
or equivalently,
\begin{eqnarray}
\nu_R(k)\sim\nu_0 \sqrt{\ln(\Lambda_0/k)}.
\label{eq:2DexponentofviscosityB}
\label{eq:2Dasympviscoink}
\end{eqnarray}
It is consistent with the result of Ref.~\cite{forster1977large} for the stirred incompressible fluids.
Substituting
Eqs.~(\ref{eq:RGincompofhc}) and (\ref{eq:2DexponentofviscosityA})
to Eq.~(\ref{eq:kappascomputationalrule}) and using $\bar\kappa_0\sim\bar \kappa_*$ [$l\to0$ of Eq.~~(\ref{eq:RGincompofhc})],
the heat conductivity is shown to satisfy  
\begin{eqnarray} 
\kappa_R (\Lambda e^{-l})\sim 
\kappa_0 \sqrt{l}/(1+\sqrt{l/l_*}),
\end{eqnarray}
or equivalently,
\begin{eqnarray} 
\kappa_R(k) \sim \kappa_0\frac{\sqrt{\ln(\Lambda_0/k)}}
{1+\sqrt{\ln(\Lambda_0/k)}/\sqrt{\ln(\Lambda_0/k_*)}},
\label{eq:2Dasympheatink}
\end{eqnarray}
where $k_*:=\Lambda_0\exp(-l_*)$.
It is the same as
\begin{eqnarray} 
\kappa_R(k) \sim \kappa_0
\sqrt{\ln[\Lambda_0/\max(k,k_*)]}
\label{eq:scalingof2Dkappainwavenumber}
\end{eqnarray}
for describing the asymptotic behaviors at $k_*\ll k\ll\Lambda_0$ and $k\ll k_*$.
Eq.~(\ref{eq:scalingof2Dkappainwavenumber}) shows that the growth of the renormalized heat conductivity begins to stop around the wavenumber $k_*$ characterizing the breakdown of the hyperscaling between the heat conductivity and kinetic viscosity in the RG flow. 

The size dependencies of the renormalized transport coefficients are satisfactorily given in the angular frequency ($\omega$) domain as $\nu_R(\omega), \kappa_R(\omega)$ rather than in the wavenumber domain, as previously known for the anomalous heat conduction~\cite{narayan2002anomalous}.
By using the asymptotic scaling exponent $z=2$ for $d=2$, the angular frequency is found to be proportional to the squared wavenumber asymptotically ($\omega\propto k^2$). This asymptotic dispersion relation rewrites the aforementioned results
Eqs.~(\ref{eq:2Dasympviscoink}) and (\ref{eq:2Dasympheatink}) in the wavenumber domain as
\begin{eqnarray} 
\nu_R(\omega)&\sim&\nu_0 \sqrt{\ln(\omega_0/\omega)}
\\
\kappa_R(\omega) &\sim& \kappa_0
\sqrt{\ln[\omega_0/\max(\omega,\omega_*)]},
\label{eq:2Dkappacutoffomega}
\end{eqnarray}
where $\omega_*:=\omega_0(k/\Lambda_0)^z$ [$\sim \omega_0( k/\Lambda_0)^2$] is the cutoff angular frequency scale, and $\omega_0$ is a value of the angular frequency giving  $\kappa_R(\omega_0)=\kappa_0$. 
Given the kinetic theory of gases~\cite{kawasaki1965logarithmic}, $\omega_0$ will be given as $\omega_0=\sqrt {Y_0}\Lambda_0$.
The same logarithmic dependencies arise in the frequency domain as in the wavenumber domain. 

The size-dependent values of observed transport coefficients $\nu_R(L)$ and $\kappa_R(L)$ at the length $L$ are given by $\nu_R(\omega)$ and $\kappa_R(\omega)$ with using
the characteristic time scale $L/c$ ($\omega=2\pi c/L$ in the angular frequency scale) for given wave speed $c$~\cite{narayan2002anomalous}; 
\begin{eqnarray} 
\nu_R(L)&\sim&\nu_0 \sqrt{\ln(L\omega_0/2\pi c)}
\\
\kappa_R(L) &\sim& \kappa_0
\sqrt{\ln[\min(L,L_*)\omega_0/(2\pi c)]},
\label{eq:2DRGkappainrealscale}
\end{eqnarray}
where $L_*:=2\pi c/\omega_*$ denotes the characteristic scale of the cutoff.

Fig.~\ref{fig:one} (a) shows an example of the numerically computed renormalized variables in the original coordinate. 
The RG flow Eq.~(\ref{eq:RGincompel}) is numerically computed, and $\nu_R$ is calculated from Eq.~(\ref{eq:nuscomputationalrule}) with $z(l)$ giving Eq.~(\ref{eq:RGincompel}); $\kappa_R/\kappa_0$ is given as the product of $\bar \kappa/\bar \kappa_0$ and $\nu_R/\nu_0$ through Eq.~(\ref{eq:kappascomputationalrule}). 
These are parametrized as functions
$\omega(l)$ given by the scaling relation Eq.~(\ref{eq:rescalingoffreq}).
Fig.~\ref{fig:one} (a) indicates the breakdown mentioned above of the hyperscaling in the angular frequency scale. 
The early divergence is shown to be slightly faster than the square-root logarithm predicted in the above asymptotic analysis and Ref.~\cite{forster1977large} because of $\mathcal O(l^{-1})$ terms neglected in these asymptotic analyses. 
Except for such a subtle difference in terms of the functional forms of divergence, the exact flow of the renormalization group is met to the above analytically obtained asymptotic behaviors of the transport coefficients. The growth of the heat conductivity is saturated, and it indicates the recovery of Fourier's law.

\subsubsection{Dynamic Renormalization-Group Analysis of One-Dimensional Materials Under Low Pressure}
With the same procedure as in the incompressible cases,
the following is obtained in the case of Eq.~(\ref{eq:1dburgwp}):
\begin{eqnarray}
&&\frac{d \nu}{dl}= \nu[z-2+\bar \lambda^2/(2\pi)]
\nonumber
\\
&&\frac{d \Sigma}{dl}= \Sigma[z-2+\bar \lambda^2/(2\pi)]
\nonumber
\\
&&\frac{d \lambda}{dl}= \lambda(z-3/2)
\label{eq:flowofburgelpassivescalarforarbitraryz}
\\
&&\frac{d Y}{dl} = Y(2z-2)
\nonumber
\\
&&\frac{d D}{dl} = D\left[ z-2+\frac{\bar \lambda ^2}{2\pi}\frac1{\bar \kappa(1+\bar\kappa)+\bar Y}
\right.\nonumber\\&&\left.
\hspace{45pt}\times\left( \frac 5 2 - \frac{\bar \kappa(1+3\bar\kappa)}{\bar \kappa(1+\bar \kappa)+\bar Y}\right)\right].
\nonumber
\end{eqnarray}
The results other than that for $\Sigma$ are the same as Ref.~\cite{PhysRevE.94.012115}.
I here corrected a mistake of Ref.~\cite{PhysRevE.94.012115} in terms of renormalized $\Sigma$, which has faultily induced the breakdown of FDR in the RG flow in the previous study.
After such correction, FDR is consequently kept naturally 
[$\Sigma(l)/\nu(l)=\Sigma_0/\nu_0$] under the RG flow of Eq.~(\ref{eq:flowofburgelpassivescalarforarbitraryz}) for an arbitrary choice of $z$.

The value of $z$ is set at $z=2-\bar\lambda^2/(2\pi)$ in order for $\nu$ to be fixed at the initial value as in the above analysis of the incompressible viscoelastic materials. The flow then reduces to
\begin{eqnarray}
&&\frac{d \bar \lambda}{dl}= \bar\lambda\left(\frac12-\frac{\bar\lambda^2}{2\pi}\right)
\nonumber
\\
&&\frac{d \bar Y}{dl} = \bar Y(2-\bar \lambda^2/\pi)
\label{eq:RGburgwp}
\\
&&\frac{d \bar\kappa}{dl} = \frac{\bar\kappa\bar \lambda^2}{2\pi}\left[ -1+\frac1{\bar \kappa(1+\bar\kappa)+\bar Y}
\right.\nonumber\\&&\left.
\hspace{45pt}\times\left( \frac 5 2 - \frac{\bar \kappa(1+3\bar\kappa)}{\bar \kappa(1+\bar \kappa)+\bar Y}\right)\right].
\nonumber
\end{eqnarray}

The reduced flow Eq.~(\ref{eq:RGburgwp}) has the following two non-trivial fixed points of giving $\bar\lambda=\sqrt{\pi} \,\,(z=3/2)$:
\begin{eqnarray}
\left(\frac 1 {(\nu\Lambda)^2}\left(\frac{\partial P}{\partial \rho}\right)_e, \frac{\kappa c_T}\nu\right)=(0,C_1(>0))
\end{eqnarray}
with 
\begin{eqnarray}
C_1&=&-\frac 2 3+\frac{1}{3\sqrt{2}} 
\nonumber
\\&&\times
\left[
\frac{{}^3\sqrt{103+9\sqrt{131}}}{2^{1/6}}
-\frac{2^{1/6}}{{}^3\sqrt{103+9\sqrt{131}}}
\right]
\end{eqnarray}
and 
\begin{eqnarray}
\left(\frac 1 {(\nu\Lambda)^2}\left(\frac{\partial P}{\partial \rho}\right)_e, \frac{\kappa c_T}\nu\right)
=
(\infty,0).
\label{eq:maybeballisticfixedpoint1D}
\end{eqnarray}
As for the case of the aforementioned incompressible viscoelastic material, 
the former is achievable only when $\bar Y_0=0$, and the latter is unconditionally linearly stable.
Note that trivial Gaussian fixed points of $\bar \lambda=0, z=2$ are unstable.
Given that the unconditionally linearly stable fixed point is uniquely that of $\bar \kappa=0$,
as long as the system has the non-zero longitudinal wave speed $\sqrt{Y}_0>0$ in the original scale, the recovery of Fourier's law is noticed to occur unavoidably in this one-dimensional model setting. 

In a similar way to that of the incompressible viscoelastic case shown earlier, the kinetic viscosity and heat conductivity are given as 
\begin{eqnarray} 
\nu_R(\omega) &\sim& \nu_0 (\omega/\omega_0)^{-1/3}
\\
\kappa_R(\omega) &\sim& \kappa_0 (\max(\omega,\omega_*)/\omega_0)^{-1/3}
\end{eqnarray}
in the angular frequency domain of the original (un-rescaled) coordinate. 
The value of $\omega_*$ is given asymptotically as $\omega_*\sim \omega_0( k_*/\Lambda_0)^{3/2}$ at $l\gg1$, which is different from $\omega_*\sim \omega_0( k_*/\Lambda_0)^2$ for the cases of the two-dimensional incompressible viscoelastic materials, due to the difference in the asymptotic values of the scaling exponent $z$ ($z=2$ for the case of the two-dimensional incompressible viscoelastic materials, $z=3/2$ for this case). 

The size-dependent values of observed transport coefficients $\nu_R(L)$ and $\kappa_R(L)$ at length $L$ are given as 
\begin{eqnarray} 
\nu_R(L) &\sim& \nu_0 [L\omega_0/(2\pi c)]^{1/3}
\\
\kappa_R(L) &\sim& \kappa_0 [\min(L,L_*)\omega_0/(2\pi c)]^{1/3}
\label{eq:RGkappainrealscale}
\end{eqnarray}
in the same way as in the two-dimensional incompressible cases.

Fig.~\ref{fig:one} (b) shows an example of the renormalized transport coefficients in the original scale predicted from the RG flow Eq.~(\ref{eq:RGburgwp}). After certain overhangs, the kinetic viscosity and heat conductivity diverge in proportion to the power of $\omega$ ($\omega^{-1/3}$) at the initial stage of the renormalization within a high angular frequency range $\omega\gg \omega_*$. At the later stage $\omega\ll\omega_*$, the hyperscaling becomes noticeably broken between the kinetic viscosity and the heat conductivity. It provides the saturation of the increase of the heat conductivity, the recovery of Fourier's law. 

Concerning the validity of the RG flow, Ref.~\cite{eyink1994renormalization} reported that the cubic- or higher-order terms such as $v^3$ are marginal for the renormalized governing equation in the case of the KPZ equation (or equivalently of the noisy Burgers equation) at $d<2$. 
The model investigated in this subsection has the same diagram expansion as that of the noisy Burgers equation at $d=1$, so the dimensional analysis provided to the KPZ equation by Ref.~\cite{eyink1994renormalization} holds for this model as well. 
Therefore, if the fluctuations are small enough at the initial condition, we can drop the third- or higher-order fluctuations throughout the calculation of the RG flow. 
Now the problem is concerning the semi-macroscopic fluctuations, and then the smallness of the fluctuation in the starting-point equation is the requirement of the equilibrium statistical mechanics. 
Given these, we are allowed to drop the third- and higher-order terms in the presented modified noisy Burgers equations. 
Note that Ref.~\cite{kardar1986dynamic} claimed that these higher orders are irrelevant, so there may be the debate whether these higher-order terms are irrelevant or marginal for the equations belonging to the KPZ class. Nonetheless, when these terms are irrelevant, we can drop them more safely, so such debates do not matter to the validity of the presented RG analysis.

\subsection{Comparison of Semi-Macroscopic Theory and Microscopic Numerics}
\subsubsection{Ballistic Scaling and Scaling Crossover in Low-Dimensional Momentum-Conserving Viscoelastic Materials}
Only the non-zero $\bar Y$ (corresponding to the square of the phase velocities) diverged under the RG flows of keeping the kinetic viscosity at a constant value. This divergence indicates that the elastic forces parametrized by $\bar Y$ dominate the long-wavelength motions of the viscoelastic materials and make the anomalous transport secondary there. 

The scaling of $z$ is thus of interest
under the flow of keeping $\bar Y$--the unique divergent term in the investigated RG flows--at a constant value, because such a scaling exponent $z$ is providing the scaling of the leading terms governing the macroscopic motions. 
Such a value of the scaling exponent $z$, setting $\bar Y$ constant, is found to be 
\begin{equation}
z=1
\end{equation}
from RG flows Eqs.~(\ref{eq:flowofincompelpassivescalarforarbitraryz}) and (\ref{eq:flowofburgelpassivescalarforarbitraryz}) for both of the investigated two models. 
The scaling given by $z=1$ (called the ballistic scaling hereafter) 
 provides the ballistic dispersion relation
 \begin{equation}
k\sim \omega.
\end{equation}
When $z=1$, the other coefficients $\nu,\Sigma,\lambda,D$ being finite or zero at $z=2,3/2$ converge to $0$ exponentially as $l$ accumulates.
It indicates that the nonlinear streaming terms and the thermal fluctuations become irrelevant for the long-wavelength motions the elastic motion dominates. 

It may be worth emphasizing that the ballistic fixed point does not mean the ballistic heat transport, which makes the energy current proportional to the temperature difference between thermal reservoirs attached to the edges of the material (not proportional to the temperature gradient within the bulk) resulting in the exponent $\alpha=1$~\cite{xiong2017crossover} apparently. Such behavior is particularly seen in the integrable systems such as the homogeneous harmonic chains~\cite{rieder1967properties}. As shown in the RG flow of the one-dimensional model with $z=3/2$ in Eq.~(\ref{eq:flowofburgelpassivescalarforarbitraryz}) (and $z=2$ in Eq.~(\ref{eq:flowofincompelpassivescalarforarbitraryz}) for $d=2$), the heat conductivity is kept finite and $\alpha=1$ is not achieved.
The name of the ballistic fixed point is just for expressing the dispersion relation realized at that fixed point.

Rather than $\alpha=1$, 
this ballistic scaling $z=1$ of the RG flows is related to 
the breakdown of the hyperscaling between the kinetic viscosity and the heat conductivity [Eqs.~(\ref{eq:maybeballisticfixedpoint2D}) and (\ref{eq:maybeballisticfixedpoint1D})], and thus to the recovery of Fourier's law,
\begin{equation}
\alpha=0.    
\end{equation}
Indeed, in Eq.~(\ref{eq:flowofburgelpassivescalarforarbitraryz}) and Eq.~(\ref{eq:flowofincompelpassivescalarforarbitraryz}) for $d=2$, $\bar \kappa$ representing the ratio of $\kappa$ to $\nu$ starts decreasing at $\bar Y\gtrsim \bar \kappa(1+\bar \kappa)$ accompanying with the explosion of $\bar Y$. 
It means that the hyperscaling is broken when the leading term becomes related to $\bar Y$. 
Then we notice that the scaling crossover from the inviscid scaling to the ballistic scaling induces the breakdown of the hypescaling, and consequently, the recovery of Fourier's law. 
This causal relationship between the ballistic scaling and recovery of Fourier's law is consistent with the one-dimensional molecular-dynamic study of Ref.~\cite{PhysRevE.94.012115} that observes $z=1$ in the spatiotemporal autocorrelation functions of diagonalized currents of mass, momentum, and energy under the thermodynamic conditions that can indicate the recovery of Fourier's law.

\subsubsection{Characteristic Length-Scale of the Scaling Crossover and the Recovery of Fourier's Law}

The above analyses suggested that the recovery of Fourier's law can be the scaling crossover between the two nontrivial fixed points of the RG flows: the inviscid and ballistic fixed points. 
It is thus interesting to infer the characteristic length-scale $L_*$, or equivalently the size-scale $N_*$ of the recovery of Fourier's law from these RG flows to verify this picture.
The following is to obtain such a relation from the RG flows to predict $L_*$ (or $N_*$), which allows us to test the developed RG analysis from the microscopic molecular-dynamic simulations. 

The anomalous heat conduction is related to the inviscid fixed point in the presented RG flows. 
Around that point, the value of $\bar Y$ is negligible, and the value of $\bar\kappa$ is given by some constant, namely,
\begin{equation}
\bar \kappa\sim C_{d}.
\label{eq:univconstinviscid}
\end{equation}
The constant $C_d$ is given by $C_{Id}$ in the flow Eq.~(\ref{eq:RGincompel}) of the incompressible viscoelastic equations, and by $C_1$ for Eq.~(\ref{eq:RGburgwp}) of the modified noisy Burgers equation of including the small pressure perturbation.

The scaling crossover has been quantified by the detachment of the flow from the inviscid fixed point.
Such detachment has attended the growth of $\bar Y$. 
The following balance condition gives the characteristic scale of starting this detachment in the investigated RG flows: 
\begin{equation}
\bar Y\sim \bar \kappa(1+\bar \kappa).
\end{equation}
This condition is satisfied at the characteristic wavenumber $k_*$.
With dimensional variables, it is rewritten as 
\begin{eqnarray}
v_{acoustic}/k_* \sim \sqrt {\kappa_* c_T(\kappa_* c_T+\nu_*)},
\label{eq:balancingconditiondimofacc}
\end{eqnarray}
where $\kappa_*$ and $\nu_*$ are the renormalized values of heat conductivity and kinetic viscosity, respectively, at the waveumber $k_*$; 
$v_{acoustic}:=\sqrt{Y}$ represents the longitudinal wave speed (that is the sound speed $c$) in one-dimensional cases, and the transverse wave speed in two-dimensional cases. 
Before the detachment is done, the flow stagnates around the inviscid fixed point, and thus $\bar \kappa$ at $l_*$ ($\kappa_*c_T/\nu_*$) is near $C_d$ of $\mathcal O(1)$ as Eq.~(\ref{eq:univconstinviscid}). 
By using it, we can rewrite the flow separation condition Eq.~(\ref{eq:balancingconditiondimofacc}) as
\begin{eqnarray}
v_{acoustic}/k_*\sim \kappa_* c_T
\label{eq:balancingconditiondim}
\end{eqnarray}
with neglecting a prefactor $\sqrt{1+1/C_d}$ of $\mathcal O(1)$.
The characteristic length $L_*=2\pi/k_*$ is then given as 
\begin{eqnarray}
L_*&\sim&  2\pi\kappa_* c_T/v_{acoustic}.
\label{eq:saturationsizeofanomalousconductionA}
\end{eqnarray} 
It gives the scaling of the recovery of Fourier's law
with the rewritten form of Eqs.~(\ref{eq:2DRGkappainrealscale}) and (\ref{eq:RGkappainrealscale}):
\begin{eqnarray}
\kappa(L) \sim \kappa_*[\min(L/L_*,1)]^\alpha.
\label{eq:wellknownformofrecov}
\end{eqnarray}
These relations Eqs.~(\ref{eq:saturationsizeofanomalousconductionA}) and (\ref{eq:wellknownformofrecov}) hold for both the one- and two-dimensional cases studied in this paper.

Besides, Eq.~(\ref{eq:numericallyfoundquantificationofGKkappa}) 
shows that the converged heat conductivity is evaluable by $\kappa_{GK}$ estimated through the Green-Kubo formula under the same ($T,P$) condition with the periodic boundary (corresponding to sufficiently large samples)~\cite{PhysRevE.94.012115}.
Eq.~(\ref{eq:numericallyfoundquantificationofGKkappa}) reduces Eqs.~(\ref{eq:saturationsizeofanomalousconductionA}) and (\ref{eq:wellknownformofrecov}) to
\begin{eqnarray}
L_*&\sim&  2\pi\kappa_{GK} c_T/v_{acoustic},
\label{eq:saturationsizeofanomalousconduction}
\\
\kappa(L) &\sim& \kappa_{GK}[\min(L/L_*,1)]^\alpha.
\label{eq:eventualroughprediction}
\end{eqnarray}
Interestingly, the variables contained in the right hand sides of Eqs.~(\ref{eq:saturationsizeofanomalousconduction}) and (\ref{eq:eventualroughprediction}) are fully evaluable in the equilibrium settings. 
Values of $c_T$ and $v_{acoustic}$ are obtained from the equilibrium statistical mechanics; the evaluation of them for the lattice systems is detailed in Ref.~\cite{spohn2014nonlinear}. The Green-Kubo formula gives $\kappa_{GK}$ from the energy-current fluctuations in the equilibrium states. 
Eqs.~(\ref{eq:saturationsizeofanomalousconduction}) and (\ref{eq:eventualroughprediction}) express a relation that relates $\kappa(N_*)$ of governing the non-equilibrium heat transport and the equilibrium fluctuations, 
and namely relates equilibrium and non-equilibrium states. 

\begin{figure}[tbp]
   \includegraphics[width=85mm]{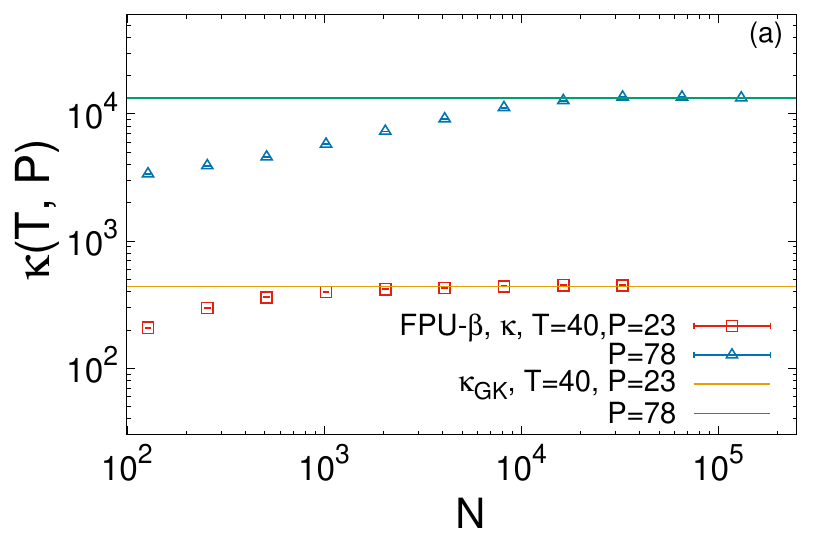}
   \includegraphics[width=85mm]{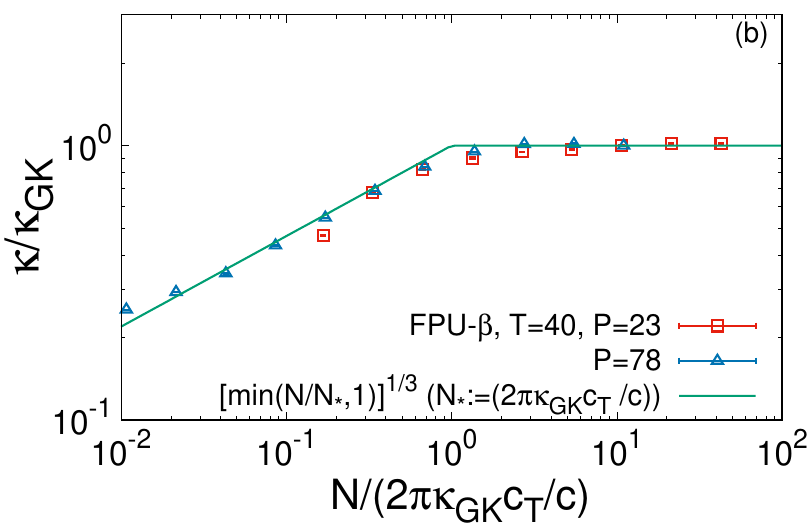}
  \caption{
Theoretical prediction of Eqs.~(\ref{eq:saturationsizeofanomalousconduction}) and (\ref{eq:eventualroughprediction}) compared with numerical experiments of one-dimensional FPU-$\beta$ lattices. 
(a) Numerical data of the heat conductivity of the FPU-$\beta$ lattices measured under non-equilibrium and equilibrium settings with given $T$ and $P$ values, after Ref.~\cite{PhysRevE.94.012115}. 
Variable $\kappa$ shows the $(T,P)$-dependent heat conductivity directly measured under steady heat conduction.
The values of $\kappa_{GK}$ express those of the heat conductivity given by the Green-Kubo formula at $N\sim10^3$ under the isolated periodic boundary condition. Temperature and pressure values are shown in legends.
(b) The scaled values of the data contained in Fig.~\ref{fig:two} (a), compared with the theoretical scaling predicted from Eqs.~(\ref{eq:saturationsizeofanomalousconduction}) and (\ref{eq:eventualroughprediction}) under conversions of $L$ to $N$ and of $v_{acoustic}$ to $c$ (the sound speed). 
  }
  \label{fig:two}
\end{figure}

The prediction of Eqs.~(\ref{eq:saturationsizeofanomalousconduction}) and (\ref{eq:eventualroughprediction}) is tested by the molecular-dynamic data of the one-dimensional FPU-$\beta$ lattices [Fig.~\ref{fig:two} (a)], where $\alpha=1/3$ has been theoretically predicted previously in the context of the anomalous heat conduction~\cite{narayan2002anomalous}. 
In the previous work~\cite{PhysRevE.94.012115}, the thermodynamic-state-dependent [$(T,P)$-dependent] heat conductivity was measured in the FPU-$\beta$ lattices of various sizes $N$ under the non-equilibrium steady states of the heat conduction. 
These size-dependent values of the heat conductivity are compared with the estimate of the heat conductivity given through the Green-Kubo formula under the periodic equilibrium conditions, which is found to be size-independent even at $N\ll N_*$.
The results 
of two parameter sets $T=40,$ $P=23$ $(22.5)$ and $T=40$, $P=78$ $(77.6)$ are shown in the figure, 
where the unit system $(m,K,\beta)=(1,1,1)$ is given by the mass $m$ of particles and the coefficients $(K,\beta)$ of the nearest-neighbor inter-particle potential $V(\Delta x)=K \Delta x^2/2+\beta \Delta x ^4/4$ depending on the  relative displacement $\Delta x$ of particles. 
Temperature and pressure are calculated from mean kinetic energy and virial theorem, respectively. 
Please refer to Ref.~\cite{PhysRevE.94.012115} for numerical details. 

Fig.~\ref{fig:two} (b) compares these numerical results
with the theoretical prediction of Eqs.~(\ref{eq:saturationsizeofanomalousconduction}) and (\ref{eq:eventualroughprediction}). 
Note that $L$ is converted to $N$ in the figure. 
The data points collapse to the predicted master curve given by Eqs.~(\ref{eq:saturationsizeofanomalousconduction}) and (\ref{eq:eventualroughprediction}) without any fitting parameters. 
This scaling crossover, occurring within longer wavelength regimes than that of the anomalous transport giving $\alpha=1/3$ theoretically predicted, is in contrast to the previously known apparent recovery of Fourier's law occurring in the intermediate regime of the ballistic one $\alpha=1$ and the anomalous one $\alpha=1/3$~\cite{chen2014nonintegrability}. 
Beyond the previous theories predicting the anomalous heat conduction,
the semi-macroscopic theory and microscopic numerics consistently suggest the new class providing the recovery of Fourier's law emerging in the low-dimensional momentum-conserving systems of the thermodynamic limit.

\section{Discussion}
The analyses of this paper have clarified that the recovery of Fourier's law can be caused by the semi-macroscopic scaling crossover in the fluctuating hydrodynamic/viscoelastic equations. There the recovery is understood as the breakdown of the hyperscaling between the viscosity and heat conductivity. 
The numerical results supported this picture through its good agreement with the predicted curve for the system-size-dependent heat conductivity, the scaling law of the recovery of Fourier's law. 

Two model equations Eqs.~(\ref{eq:RGincompel}) and (\ref{eq:RGburgwp}) have demonstrated the possible universality of the recovery of Fourier's law in low-dimensional momentum-conserving viscoelastic systems.
The intrinsic feature involved in the presented calculation is associated with
the dimension-independent change 
$G_0=i\omega+\nu_0k^2 \to i\omega + [\nu_0+Y_0/(i\omega)]k^2$ 
of the velocity field Green's function in both models.
The pole ($\omega=i\nu_0k^2$) of that Green's function for the fluids deviates from the imaginary axis in the frequency space due to the non-zero wave propagation speeds ($\sqrt Y_0>0$) in the viscoelastic materials. 
It consequently kept the renormalized heat conductivity finite in the presented analysis.
In the sense that the long-range interactions truncate the massless behavior of systems, we may speculate that this mechanism of the recovery of Fourier's law has a somewhat similar mathematical structure to the Anderson-Higgs mechanisms of the superconductivity in electromagnetism~\cite{aitchison2012gauge}.
Quite recent report~\cite{di2019equilibrium} of the recovery of Fourier's law in many-particle systems of assuming long-range microscopic interactions may be supportive about this speculation. 

The presented one- and two-dimensional analyses will consistently explain the apparently-conflicting relationship between the two-dimensional incompressible fluids showing the anomaly~\cite{forster1977large} and 
one-dimensional fluids with pressure showing the recovery~\cite{PhysRevE.94.012115}. 
In one-dimensional cases, the anomalous heat conduction seems unstable against the perturbation of the thermal pressure fluctuations, occurring in both fluids and viscoelastic materials (or rather, there is no distinction between them). 
The two-dimensional case is in contrast to it because the recovery of Fourier's law requires the non-zero elastic shear stress to inhibit the anomalous heat conduction.
Namely, the existence of the pressure is not a sufficient condition to yield the recovery of Fourier's law in $d>1$. 
These suggest that the recovery of Fourier's law is a universal phenomenon for low-dimensional solid systems. 

Experimentally available low-dimensional systems are frequently the solid materials, such as the carbon nanotube and graphene sheets, so this recovery of Fourier's law will be an experimentally accessible phenomenon. These novel materials accelerate the recent experimental research of the low-dimensional transports~\cite{chang2008breakdown,balandin2011thermal}.
In that context, the experimental test for the recovery of Fourier's law would be intriguing as it is related to a sort of the non-equilibrium universal scaling Eq.~(\ref{eq:eventualroughprediction}). It would also be of interest to verify whether such a non-equilibrium scaling links to the equilibrium fluctuations as predicted by Eq.~(\ref{eq:saturationsizeofanomalousconduction}).

There had been at least two distinct origins of 
the recovery of Fourier's law [Eq.~(\ref{eq:predprevstud})] in the low-dimensional momentum-conserving many-particle systems~\cite{PhysRevE.94.012115}. 
One is the thermally activated dissociation~\cite{gendelman2014normal}. 
Another has now been clarified to be the scaling crossover in the semi-macroscopic viscoelastic materials. 
We can distinguish them through the temperature-dependence of the heat conductivity~\cite{PhysRevE.94.012115}. Dissociation-induced recovery gives the heat conductivity the temperature dependence of which is consistent with the Arrhenius equation~\cite{gendelman2014normal}, and the scaling-crossover-induced one shows another form of the temperature dependence~\cite{PhysRevE.94.012115}. 
One-dimensional soft-core systems are the typical examples of the dissociation-induced recovery~\cite{gendelman2014normal,PhysRevE.94.012115}; please refer to Ref.~\cite{PhysRevE.94.012115} for the estimate of $L_*$ ($N_*$) in the cases of the dissociation-induced recovery. 
The heat conductivity of the FPU-$\beta$ model showed the power-law dependence on the temperature~\cite{aoki2001fermi,PhysRevE.94.012115}, and as seen in Fig.~\ref{fig:two}, the FPU-$\beta$ model is a typical example for the scaling-crossover-induced recovery of Fourier's law obeying Eq.~(\ref{eq:saturationsizeofanomalousconduction}).  
Meanwhile, the requirements to cause these two routes of the recovery of Fourier's law are still uncertain. The dissociation is for inhibiting the semi-macroscopic collective motions, so the semi-macroscopic analysis (assuming such collective motions) cannot predict the rate-theoretic dissociation. Further simulations will be needed to overcome this difficulty and to clarify the conditions for involving these two kinds of the recovery of Fourier's law. 
Although the above classifications have been investigated mostly in one-dimensional systems, the two-dimensional systems, being recently reported to show the recovery of Fourier's law~\cite{savin2016normal}, will also have such systematics. 

The presented explicit forms of the RG flows allow us to locate the recovery of Fourier's law in the context of the anomalous heat conduction.
Refs.~\cite{narayan2002anomalous,mai2006universality} provided the general proof for the existence of the nontrivial fixed point showing the anomalous heat conduction in one-dimensional fluctuating hydrodynamic/viscoelastic equations. Such a fixed point is indeed observed in the RG flows investigated in this study as well.
However, to conclude the anomalous heat conduction from the existence of that fixed point, we need the stability of it in the RG flows. 
The scope of Refs.~\cite{narayan2002anomalous,mai2006universality} was to show the persistent existence of the fixed point of the anomalous heat conduction, 
and they did not mention its stability.  
Note that the Galilean invariance of the model equation (corresponding to $\lambda=1$), used to obtain the inviscid fixed point~\cite{narayan2002anomalous}, is not the necessary condition in the RG flow~\cite{eyink1994renormalization}; its obvious example is the Gaussian fixed point at which $\bar\lambda$ is zero~\cite{eyink1994renormalization}. 
As shown in this study, 
the fixed point of the anomalous heat conduction can be unstable due to the elastic response of the materials, and thus the recovery of Fourier's law can emerge without any conflict with the properties of the RG flows proved by Refs.~\cite{narayan2002anomalous,mai2006universality}. 
Another study \cite{spohn2016fluctuating} obtained the asymptotic autocorrelation functions of the currents in the fluctuating hydrodynamic equations, with assuming the self-similar scaling of $z=3/2$ of the anomalous heat conduction. 
As shown in this study, however, the scaling of leading-order motions can deviate from $z=3/2$ to $z=1$ when the recovery of Fourier's law emerges. 
Such scaling crossover has been indeed numerically detected with the FPU-$\beta$ lattices in Ref.~\cite{PhysRevE.94.012115}, where the scaling crossover from $z=3/2$ to $z=1$ was observed in the current autocorrelation functions as the volumetric strain (and pressure) increased to certain values resulting in the recovery of Fourier's law. 
Ref.~\cite{spohn2016fluctuating} is as above the theory for the anomalous heat conduction utilizing the scaling exponent $z=3/2$, 
and so the existence of the recovery of Fourier's law, characterized by the exponent $z=1$, is not conflicting with it.

It may be interesting 
that the renormalization group  [Fig.~\ref{fig:one} (b)] 
captures 
an apparent recovery of Fourier's law at quite high angular frequencies $\omega\sim\omega_0$, as well as the true recovery of Fourier's law at $\omega<\omega_*$ involving the thermodynamic limit. 
The apparent plateau $\kappa\sim\kappa_0$ at $\omega\sim\omega_0$ perhaps corresponds to the numerically-detected transient recovery of Fourier's law~\cite{chen2014nonintegrability} occurring within the intermediate scale between the size scales of the ballistic transport $\alpha=1$ and anomalous transport $\alpha=1/3$; such a transient plateau is observed in the molecular-dynamic simulations even after carefully considering the interfacial thermal resistance~\cite{PhysRevE.94.012115}. 
Numerically-measurable apparently-convergent heat conductivity may be indicating the bare parameter $\kappa_0$ of the heat conduction. The data of the FPU-$\beta$ lattices in Fig.~\ref{fig:two} at $P=78, N\sim10^2$ is suggestive. 
The investigation of the bare parameter will be of interest in the fundamental studies of the non-equilibrium statistical mechanics.

Under the linearization of the elastic forces, this study clarified that the semi-macroscopic RG flow changes even around the inviscid fixed point (of negligible elastic effects) due to the elastic effects. 
The assumptions and approximations of this study are, as mentioned earlier, those of previous studies and such linearization dropping the square-orders proportional to the elastic coefficients. 
The approximations of the previous studies have been verified and examined theoretically~\cite{forster1977large,kardar1986dynamic,eyink1994renormalization,spohn2014nonlinear}, so the fundamental assumption of this study will be the phenomenologically-introduced linearization of the elastic effects; the verification on the governing equation is detailed in Ref.~\cite{spohn2014nonlinear} (also please refer to \S\ref{sec:modelintroduction}), and the irrelevance or marginal properties of the cubic orders in the RG flows around the inviscid point has been shown in Refs.~\cite{forster1977large,kardar1986dynamic,eyink1994renormalization}. 
Indeed, for example, even if we keep the nonlinear terms $P v_a$ of the non-dissipative momentum currents without using the decoupling hypothesis, we can obtain the same diagram calculation for the renormalized transport coefficients~\cite{PhysRevE.94.012115}. 
Regarding the linearlized elastic effects, the elastic terms are infinitesimal around the previously-obtained inviscid fixed point of the anomalous heat conduction, so the linearization of the elastic forces becomes exact there. 
Given these, the elasticity-induced instability (pressure-induced in one-dimensional cases and rigidity-induced in two-dimensional cases) of that fixed point will be a very robust result of this study. 
The elasticity yields a relevant parameter of the RG flow, the longitudinal- or transverse-wave speed, not discussed in the previous studies as above. 

Compared to the clarification of the relevance of the elastic effects in the RG flows, the presented RG flows far from the inviscid fixed point are phenomenological. 
They will be modified when we investigate the full hydrodynamic/viscoelastic equations. 
Indeed, the RG analyses of the addressed one-dimensional model in this study cannot explain why the anomalous heat conduction is sustained under zero-pressure 
while the recovery of Fourier's law occurs at finite pressure in the previous simulations of FPU-$\beta$ lattices~\cite{PhysRevE.94.012115}. 
Regrettably, the inviscid fixed point of the anomalous heat conduction is always unconditionally unstable in the model equations of this study assuming linearizable small elastic properties. 
The full-analysis of considering non-linear elastic effects will clarify the correct phase transition point between the anomalous heat conduction and the recovery of Fourier's law. 
Nevertheless, like the previous model~\cite{forster1977large} gives the exact renormalized equations for the small thermal fluctuations around the fixed points (the linear diffusion equation around $z=2$ and the diffusion equation with streaming terms around $z=1+d/2$~\cite{narayan2002anomalous}), the presented models provide the exact renormalized equations around $z=1$ (the wave equation) as well as around $z=2$ and $z=1+d/2$. 
The presented models then successfully capture the minimal governing equations dominating these fixed points.
Given that point, although the presented model equations cannot predict the accurate transition point, 
the recovery of Fourier's law will exist in the full-analysis, as far as it can be interpreted from the scaling crossover between the inviscid fixed point and ballistic fixed point. Such a transition is thus currently a conjecture, yet being strongly supported from the molecular-dynamic viewpoint (Fig.~\ref{fig:two}).

Most of the validity of the theoretical prediction of this study far from the inviscid fixed point is as above currently based on the numerical supports shown in Fig.~\ref{fig:two} and in Ref.~\cite{PhysRevE.94.012115}. 
Although we can obtain the ballistic scaling from the full fluctuating viscoelastic/hydrodynamic equations with requiring the balance between the mass and momentum density fluctuations under the invariance of the mass-conservation law~\cite{PhysRevE.94.012115} as we do the inviscid fixed point with requiring the balance between the energy density and velocity fluctuations under the Galilean invariance~\cite{narayan2002anomalous}, such discussion is just to show the existence of the fixed points as mentioned earlier. 
Obtaining the RG flow of the full-order viscoelastic equations, which will give the decisive conclusion, is a hard task, so the molecular-dynamic study will play an essential role to access this dynamic non-equilibrium phase transition. 
Given the transition in the RG flows is based on the semi-macroscopic thermodynamics, the transition point should be written by the thermodynamic quantities, involving the semi-macroscopic order parameters and the transport coefficients perhaps. 
In the case of FPU-$\beta$ lattices, the transition is observed near the conditions of unit inter-particle strain $l\sim 1$ ($P\sim 10^1 $) in the temperature range $T\in [1, 50]$ (under the same normalization as in Fig.~\ref{fig:two})~\cite{PhysRevE.94.012115}. 
Moderately-high temperature is required at least to get the non-integrable/non-linear natures of the microscopic potential energy (corresponding to the phonon scattering); that makes the temperature gradient be the dominant thermodynamic force~\cite{casati1984one,chen2014nonintegrability} whereas the interfacial temperature gap is dominant in integrable systems~\cite{rieder1967properties}. 
The monatomic lattices of nearest-neighbor-potentials will be the closest systems in accessing this issue since the other particle systems (non-monaotmic, non-nearest-neighbor-potential, or non-momentum-conserving systems) are somehow related to other issues as below. 
The pinning potential should be zero to obtain the momentum-conservation; if not, the hydrodynamic anomaly is originally not existing in the many-particle systems, and there will be the normal heat conduction under sufficient non-linearity~\cite{casati1984one}. Regularly-ordered particle mass is required to avoid the insulator phase realized in the random-mass systems~\cite{matsuda1970localization,dhar2001heat}. Besides, diatomic/multi-atomic systems may have another order as the multi-phase fluids~\cite{PhysRevE.94.012115}. 
The long-range inter-particle potential~\cite{di2019equilibrium} will produce another order like electromagnetic fields. It is also of interest whether these systems still show the phase-transition treated in this study, yet these systems may lead to more comprehensive taxonomy of the heat conduction in the low-dimensional lattices.

\section{Conclusion}
In order to clarify the mechanism of the recovery of Fourier's law in the low-dimensional momentum-conserving systems reported by previous molecular-dynamic simulations,
the dynamic renormalization-group analyses are presented for two simplified model cases of the fluctuating viscoelastic equations. 
One model is incompressible one widely known to show the divergence of the transport coefficients for fluidic cases. The other is the modified noisy Burgers fluid containing small pressure perturbation, known originally to show the anomalous transport at zero pressure.
The recovery of Fourier's law is shown to emerge as a scaling crossover in the investigated RG flows. 
The predicted scaling quantitatively coincided with the heat conductivity observed in the molecular dynamics of the FPU-$\beta$ lattices. 
These results consistently suggested that the recovery of Fourier's law can provide a universality class to the low-dimensional momentum-conserving solid systems.

\begin{acknowledgements}
The author gratefully acknowledges helpful discussions with H. Hayakawa and S. Takesue.
The author gratefully thanks T. Hatano and anonymous reviewers who patiently helped improve the manuscript.
\end{acknowledgements}

\begin{thebibliography}{47}%
\makeatletter
\providecommand \@ifxundefined [1]{%
 \@ifx{#1\undefined}
}%
\providecommand \@ifnum [1]{%
 \ifnum #1\expandafter \@firstoftwo
 \else \expandafter \@secondoftwo
 \fi
}%
\providecommand \@ifx [1]{%
 \ifx #1\expandafter \@firstoftwo
 \else \expandafter \@secondoftwo
 \fi
}%
\providecommand \natexlab [1]{#1}%
\providecommand \enquote  [1]{``#1''}%
\providecommand \bibnamefont  [1]{#1}%
\providecommand \bibfnamefont [1]{#1}%
\providecommand \citenamefont [1]{#1}%
\providecommand \href@noop [0]{\@secondoftwo}%
\providecommand \href [0]{\begingroup \@sanitize@url \@href}%
\providecommand \@href[1]{\@@startlink{#1}\@@href}%
\providecommand \@@href[1]{\endgroup#1\@@endlink}%
\providecommand \@sanitize@url [0]{\catcode `\\12\catcode `\$12\catcode
  `\&12\catcode `\#12\catcode `\^12\catcode `\_12\catcode `\%12\relax}%
\providecommand \@@startlink[1]{}%
\providecommand \@@endlink[0]{}%
\providecommand \url  [0]{\begingroup\@sanitize@url \@url }%
\providecommand \@url [1]{\endgroup\@href {#1}{\urlprefix }}%
\providecommand \urlprefix  [0]{URL }%
\providecommand \Eprint [0]{\href }%
\providecommand \doibase [0]{http://dx.doi.org/}%
\providecommand \selectlanguage [0]{\@gobble}%
\providecommand \bibinfo  [0]{\@secondoftwo}%
\providecommand \bibfield  [0]{\@secondoftwo}%
\providecommand \translation [1]{[#1]}%
\providecommand \BibitemOpen [0]{}%
\providecommand \bibitemStop [0]{}%
\providecommand \bibitemNoStop [0]{.\EOS\space}%
\providecommand \EOS [0]{\spacefactor3000\relax}%
\providecommand \BibitemShut  [1]{\csname bibitem#1\endcsname}%
\let\auto@bib@innerbib\@empty
\bibitem [{\citenamefont {De~Groot}\ and\ \citenamefont
  {Mazur}(2013)}]{de2013non}%
  \BibitemOpen
  \bibfield  {author} {\bibinfo {author} {\bibfnamefont {S.~R.}\ \bibnamefont
  {De~Groot}}\ and\ \bibinfo {author} {\bibfnamefont {P.}~\bibnamefont
  {Mazur}},\ }\href@noop {} {\emph {\bibinfo {title} {Non-equilibrium
  thermodynamics}}}\ (\bibinfo  {publisher} {Courier Corporation},\ \bibinfo
  {year} {2013})\BibitemShut {NoStop}%
\bibitem [{\citenamefont {Seifert}(2012)}]{seifert2012stochastic}%
  \BibitemOpen
  \bibfield  {author} {\bibinfo {author} {\bibfnamefont {U.}~\bibnamefont
  {Seifert}},\ }\href@noop {} {\bibfield  {journal} {\bibinfo  {journal}
  {Reports on Progress in Physics}\ }\textbf {\bibinfo {volume} {75}},\
  \bibinfo {pages} {126001} (\bibinfo {year} {2012})}\BibitemShut {NoStop}%
\bibitem [{\citenamefont {Casati}\ \emph {et~al.}(1984)\citenamefont {Casati},
  \citenamefont {Ford}, \citenamefont {Vivaldi},\ and\ \citenamefont
  {Visscher}}]{casati1984one}%
  \BibitemOpen
  \bibfield  {author} {\bibinfo {author} {\bibfnamefont {G.}~\bibnamefont
  {Casati}}, \bibinfo {author} {\bibfnamefont {J.}~\bibnamefont {Ford}},
  \bibinfo {author} {\bibfnamefont {F.}~\bibnamefont {Vivaldi}}, \ and\
  \bibinfo {author} {\bibfnamefont {W.~M.}\ \bibnamefont {Visscher}},\
  }\href@noop {} {\bibfield  {journal} {\bibinfo  {journal} {Physical review
  letters}\ }\textbf {\bibinfo {volume} {52}},\ \bibinfo {pages} {1861}
  (\bibinfo {year} {1984})}\BibitemShut {NoStop}%
\bibitem [{\citenamefont {Takesue}(1990)}]{takesue1990fourier}%
  \BibitemOpen
  \bibfield  {author} {\bibinfo {author} {\bibfnamefont {S.}~\bibnamefont
  {Takesue}},\ }\href@noop {} {\bibfield  {journal} {\bibinfo  {journal}
  {Physical review letters}\ }\textbf {\bibinfo {volume} {64}},\ \bibinfo
  {pages} {252} (\bibinfo {year} {1990})}\BibitemShut {NoStop}%
\bibitem [{\citenamefont {Balandin}(2011)}]{balandin2011thermal}%
  \BibitemOpen
  \bibfield  {author} {\bibinfo {author} {\bibfnamefont {A.~A.}\ \bibnamefont
  {Balandin}},\ }\href@noop {} {\bibfield  {journal} {\bibinfo  {journal}
  {Nature materials}\ }\textbf {\bibinfo {volume} {10}},\ \bibinfo {pages}
  {569} (\bibinfo {year} {2011})}\BibitemShut {NoStop}%
\bibitem [{\citenamefont {Gu}\ \emph {et~al.}(2018)\citenamefont {Gu},
  \citenamefont {Wei}, \citenamefont {Yin}, \citenamefont {Li},\ and\
  \citenamefont {Yang}}]{gu2018colloquium}%
  \BibitemOpen
  \bibfield  {author} {\bibinfo {author} {\bibfnamefont {X.}~\bibnamefont
  {Gu}}, \bibinfo {author} {\bibfnamefont {Y.}~\bibnamefont {Wei}}, \bibinfo
  {author} {\bibfnamefont {X.}~\bibnamefont {Yin}}, \bibinfo {author}
  {\bibfnamefont {B.}~\bibnamefont {Li}}, \ and\ \bibinfo {author}
  {\bibfnamefont {R.}~\bibnamefont {Yang}},\ }\href@noop {} {\bibfield
  {journal} {\bibinfo  {journal} {Reviews of Modern Physics}\ }\textbf
  {\bibinfo {volume} {90}},\ \bibinfo {pages} {041002} (\bibinfo {year}
  {2018})}\BibitemShut {NoStop}%
\bibitem [{\citenamefont {Forster}\ \emph {et~al.}(1977)\citenamefont
  {Forster}, \citenamefont {Nelson},\ and\ \citenamefont
  {Stephen}}]{forster1977large}%
  \BibitemOpen
  \bibfield  {author} {\bibinfo {author} {\bibfnamefont {D.}~\bibnamefont
  {Forster}}, \bibinfo {author} {\bibfnamefont {D.~R.}\ \bibnamefont {Nelson}},
  \ and\ \bibinfo {author} {\bibfnamefont {M.~J.}\ \bibnamefont {Stephen}},\
  }\href@noop {} {\bibfield  {journal} {\bibinfo  {journal} {Physical Review
  A}\ }\textbf {\bibinfo {volume} {16}},\ \bibinfo {pages} {732} (\bibinfo
  {year} {1977})}\BibitemShut {NoStop}%
\bibitem [{\citenamefont {Faris}\ and\ \citenamefont
  {Jona-Lasinio}(1982)}]{faris1982large}%
  \BibitemOpen
  \bibfield  {author} {\bibinfo {author} {\bibfnamefont {W.~G.}\ \bibnamefont
  {Faris}}\ and\ \bibinfo {author} {\bibfnamefont {G.}~\bibnamefont
  {Jona-Lasinio}},\ }\href@noop {} {\bibfield  {journal} {\bibinfo  {journal}
  {Journal of Physics A: Mathematical and General}\ }\textbf {\bibinfo {volume}
  {15}},\ \bibinfo {pages} {3025} (\bibinfo {year} {1982})}\BibitemShut
  {NoStop}%
\bibitem [{\citenamefont {Lepri}\ \emph {et~al.}(2003)\citenamefont {Lepri},
  \citenamefont {Livi},\ and\ \citenamefont {Politi}}]{lepri2003thermal}%
  \BibitemOpen
  \bibfield  {author} {\bibinfo {author} {\bibfnamefont {S.}~\bibnamefont
  {Lepri}}, \bibinfo {author} {\bibfnamefont {R.}~\bibnamefont {Livi}}, \ and\
  \bibinfo {author} {\bibfnamefont {A.}~\bibnamefont {Politi}},\ }\href@noop {}
  {\bibfield  {journal} {\bibinfo  {journal} {Physics Reports}\ }\textbf
  {\bibinfo {volume} {377}},\ \bibinfo {pages} {1} (\bibinfo {year}
  {2003})}\BibitemShut {NoStop}%
\bibitem [{\citenamefont {Saito}\ and\ \citenamefont
  {Dhar}(2007)}]{saito2007fluctuation}%
  \BibitemOpen
  \bibfield  {author} {\bibinfo {author} {\bibfnamefont {K.}~\bibnamefont
  {Saito}}\ and\ \bibinfo {author} {\bibfnamefont {A.}~\bibnamefont {Dhar}},\
  }\href@noop {} {\bibfield  {journal} {\bibinfo  {journal} {Physical Review
  Letters}\ }\textbf {\bibinfo {volume} {99}},\ \bibinfo {pages} {180601}
  (\bibinfo {year} {2007})}\BibitemShut {NoStop}%
\bibitem [{\citenamefont {Kawasaki}\ and\ \citenamefont
  {Oppenheim}(1965)}]{kawasaki1965logarithmic}%
  \BibitemOpen
  \bibfield  {author} {\bibinfo {author} {\bibfnamefont {K.}~\bibnamefont
  {Kawasaki}}\ and\ \bibinfo {author} {\bibfnamefont {I.}~\bibnamefont
  {Oppenheim}},\ }\href@noop {} {\bibfield  {journal} {\bibinfo  {journal}
  {Physical Review}\ }\textbf {\bibinfo {volume} {139}},\ \bibinfo {pages}
  {A1763} (\bibinfo {year} {1965})}\BibitemShut {NoStop}%
\bibitem [{\citenamefont {Pomeau}\ and\ \citenamefont
  {Resibois}(1975)}]{pomeau1975time}%
  \BibitemOpen
  \bibfield  {author} {\bibinfo {author} {\bibfnamefont {Y.}~\bibnamefont
  {Pomeau}}\ and\ \bibinfo {author} {\bibfnamefont {P.}~\bibnamefont
  {Resibois}},\ }\href@noop {} {\bibfield  {journal} {\bibinfo  {journal}
  {Physics Reports}\ }\textbf {\bibinfo {volume} {19}},\ \bibinfo {pages} {63}
  (\bibinfo {year} {1975})}\BibitemShut {NoStop}%
\bibitem [{\citenamefont {Lepri}\ \emph {et~al.}(1998)\citenamefont {Lepri},
  \citenamefont {Livi},\ and\ \citenamefont {Politi}}]{lepri1998anomalous}%
  \BibitemOpen
  \bibfield  {author} {\bibinfo {author} {\bibfnamefont {S.}~\bibnamefont
  {Lepri}}, \bibinfo {author} {\bibfnamefont {R.}~\bibnamefont {Livi}}, \ and\
  \bibinfo {author} {\bibfnamefont {A.}~\bibnamefont {Politi}},\ }\href@noop {}
  {\bibfield  {journal} {\bibinfo  {journal} {EPL (Europhysics Letters)}\
  }\textbf {\bibinfo {volume} {43}},\ \bibinfo {pages} {271} (\bibinfo {year}
  {1998})}\BibitemShut {NoStop}%
\bibitem [{\citenamefont {Chang}\ \emph {et~al.}(2008)\citenamefont {Chang},
  \citenamefont {Okawa}, \citenamefont {Garcia}, \citenamefont {Majumdar},\
  and\ \citenamefont {Zettl}}]{chang2008breakdown}%
  \BibitemOpen
  \bibfield  {author} {\bibinfo {author} {\bibfnamefont {C.-W.}\ \bibnamefont
  {Chang}}, \bibinfo {author} {\bibfnamefont {D.}~\bibnamefont {Okawa}},
  \bibinfo {author} {\bibfnamefont {H.}~\bibnamefont {Garcia}}, \bibinfo
  {author} {\bibfnamefont {A.}~\bibnamefont {Majumdar}}, \ and\ \bibinfo
  {author} {\bibfnamefont {A.}~\bibnamefont {Zettl}},\ }\href@noop {}
  {\bibfield  {journal} {\bibinfo  {journal} {Physical review letters}\
  }\textbf {\bibinfo {volume} {101}},\ \bibinfo {pages} {075903} (\bibinfo
  {year} {2008})}\BibitemShut {NoStop}%
\bibitem [{\citenamefont {Lepri}\ \emph {et~al.}(2016)\citenamefont {Lepri},
  \citenamefont {Livi},\ and\ \citenamefont {Politi}}]{lepri2016heat}%
  \BibitemOpen
  \bibfield  {author} {\bibinfo {author} {\bibfnamefont {S.}~\bibnamefont
  {Lepri}}, \bibinfo {author} {\bibfnamefont {R.}~\bibnamefont {Livi}}, \ and\
  \bibinfo {author} {\bibfnamefont {A.}~\bibnamefont {Politi}},\ }in\
  \href@noop {} {\emph {\bibinfo {booktitle} {Thermal Transport in Low
  Dimensions}}}\ (\bibinfo  {publisher} {Springer},\ \bibinfo {year} {2016})\
  pp.\ \bibinfo {pages} {1--37}\BibitemShut {NoStop}%
\bibitem [{\citenamefont {Landau}\ and\ \citenamefont
  {Lifshitz}(1986)}]{landau1986course}%
  \BibitemOpen
  \bibfield  {author} {\bibinfo {author} {\bibfnamefont {L.~D.}\ \bibnamefont
  {Landau}}\ and\ \bibinfo {author} {\bibfnamefont {E.~M.}\ \bibnamefont
  {Lifshitz}},\ }\href@noop {} {\emph {\bibinfo {title} {Course of theoretical
  physics, Theory of elasticity}}}\ (\bibinfo  {publisher} {Pergamon Press},\
  \bibinfo {year} {1986})\BibitemShut {NoStop}%
\bibitem [{\citenamefont {Narayan}\ and\ \citenamefont
  {Ramaswamy}(2002)}]{narayan2002anomalous}%
  \BibitemOpen
  \bibfield  {author} {\bibinfo {author} {\bibfnamefont {O.}~\bibnamefont
  {Narayan}}\ and\ \bibinfo {author} {\bibfnamefont {S.}~\bibnamefont
  {Ramaswamy}},\ }\href@noop {} {\bibfield  {journal} {\bibinfo  {journal}
  {Physical review letters}\ }\textbf {\bibinfo {volume} {89}},\ \bibinfo
  {pages} {200601\_1} (\bibinfo {year} {2002})}\BibitemShut {NoStop}%
\bibitem [{\citenamefont {Spohn}(2016)}]{spohn2016fluctuating}%
  \BibitemOpen
  \bibfield  {author} {\bibinfo {author} {\bibfnamefont {H.}~\bibnamefont
  {Spohn}},\ }in\ \href@noop {} {\emph {\bibinfo {booktitle} {Thermal Transport
  in Low Dimensions}}}\ (\bibinfo  {publisher} {Springer},\ \bibinfo {year}
  {2016})\ pp.\ \bibinfo {pages} {107--158}\BibitemShut {NoStop}%
\bibitem [{\citenamefont {Gendelman}\ and\ \citenamefont
  {Savin}(2000)}]{gendelman2000normal}%
  \BibitemOpen
  \bibfield  {author} {\bibinfo {author} {\bibfnamefont {O.~V.}\ \bibnamefont
  {Gendelman}}\ and\ \bibinfo {author} {\bibfnamefont {A.~V.}\ \bibnamefont
  {Savin}},\ }\href@noop {} {\bibfield  {journal} {\bibinfo  {journal}
  {Physical review letters}\ }\textbf {\bibinfo {volume} {84}},\ \bibinfo
  {pages} {2381} (\bibinfo {year} {2000})}\BibitemShut {NoStop}%
\bibitem [{\citenamefont {Jiang}\ and\ \citenamefont
  {Zhao}(2016)}]{jiang2016modulating}%
  \BibitemOpen
  \bibfield  {author} {\bibinfo {author} {\bibfnamefont {J.}~\bibnamefont
  {Jiang}}\ and\ \bibinfo {author} {\bibfnamefont {H.}~\bibnamefont {Zhao}},\
  }\href@noop {} {\bibfield  {journal} {\bibinfo  {journal} {Journal of
  Statistical Mechanics: Theory and Experiment}\ }\textbf {\bibinfo {volume}
  {2016}},\ \bibinfo {pages} {093208} (\bibinfo {year} {2016})}\BibitemShut
  {NoStop}%
\bibitem [{\citenamefont {Sato}(2016)}]{PhysRevE.94.012115}%
  \BibitemOpen
  \bibfield  {author} {\bibinfo {author} {\bibfnamefont {D.~S.~K.}\
  \bibnamefont {Sato}},\ }\href {\doibase 10.1103/PhysRevE.94.012115}
  {\bibfield  {journal} {\bibinfo  {journal} {Phys. Rev. E}\ }\textbf {\bibinfo
  {volume} {94}},\ \bibinfo {pages} {012115} (\bibinfo {year}
  {2016})}\BibitemShut {NoStop}%
\bibitem [{\citenamefont {Zhong}\ \emph {et~al.}(2012)\citenamefont {Zhong},
  \citenamefont {Zhang}, \citenamefont {Wang},\ and\ \citenamefont
  {Zhao}}]{zhong2012normal}%
  \BibitemOpen
  \bibfield  {author} {\bibinfo {author} {\bibfnamefont {Y.}~\bibnamefont
  {Zhong}}, \bibinfo {author} {\bibfnamefont {Y.}~\bibnamefont {Zhang}},
  \bibinfo {author} {\bibfnamefont {J.}~\bibnamefont {Wang}}, \ and\ \bibinfo
  {author} {\bibfnamefont {H.}~\bibnamefont {Zhao}},\ }\href@noop {} {\bibfield
   {journal} {\bibinfo  {journal} {Physical Review E}\ }\textbf {\bibinfo
  {volume} {85}},\ \bibinfo {pages} {060102(R)} (\bibinfo {year}
  {2012})}\BibitemShut {NoStop}%
\bibitem [{\citenamefont {Chen}\ \emph {et~al.}(2012)\citenamefont {Chen},
  \citenamefont {Zhang}, \citenamefont {Wang},\ and\ \citenamefont
  {Zhao}}]{chen2012breakdown}%
  \BibitemOpen
  \bibfield  {author} {\bibinfo {author} {\bibfnamefont {S.}~\bibnamefont
  {Chen}}, \bibinfo {author} {\bibfnamefont {Y.}~\bibnamefont {Zhang}},
  \bibinfo {author} {\bibfnamefont {J.}~\bibnamefont {Wang}}, \ and\ \bibinfo
  {author} {\bibfnamefont {H.}~\bibnamefont {Zhao}},\ }\href@noop {} {\bibfield
   {journal} {\bibinfo  {journal} {arXiv preprint arXiv:1204.5933}\ } (\bibinfo
  {year} {2012})}\BibitemShut {NoStop}%
\bibitem [{\citenamefont {Das}\ \emph {et~al.}(2014)\citenamefont {Das},
  \citenamefont {Dhar},\ and\ \citenamefont {Narayan}}]{das2014heat}%
  \BibitemOpen
  \bibfield  {author} {\bibinfo {author} {\bibfnamefont {S.~G.}\ \bibnamefont
  {Das}}, \bibinfo {author} {\bibfnamefont {A.}~\bibnamefont {Dhar}}, \ and\
  \bibinfo {author} {\bibfnamefont {O.}~\bibnamefont {Narayan}},\ }\href@noop
  {} {\bibfield  {journal} {\bibinfo  {journal} {Journal of Statistical
  Physics}\ }\textbf {\bibinfo {volume} {154}},\ \bibinfo {pages} {204}
  (\bibinfo {year} {2014})}\BibitemShut {NoStop}%
\bibitem [{\citenamefont {Chen}\ \emph {et~al.}(2014)\citenamefont {Chen},
  \citenamefont {Wang}, \citenamefont {Casati},\ and\ \citenamefont
  {Benenti}}]{chen2014nonintegrability}%
  \BibitemOpen
  \bibfield  {author} {\bibinfo {author} {\bibfnamefont {S.}~\bibnamefont
  {Chen}}, \bibinfo {author} {\bibfnamefont {J.}~\bibnamefont {Wang}}, \bibinfo
  {author} {\bibfnamefont {G.}~\bibnamefont {Casati}}, \ and\ \bibinfo {author}
  {\bibfnamefont {G.}~\bibnamefont {Benenti}},\ }\href@noop {} {\bibfield
  {journal} {\bibinfo  {journal} {Physical Review E}\ }\textbf {\bibinfo
  {volume} {90}},\ \bibinfo {pages} {032134} (\bibinfo {year}
  {2014})}\BibitemShut {NoStop}%
\bibitem [{\citenamefont {Kubo}\ \emph {et~al.}(2012)\citenamefont {Kubo},
  \citenamefont {Toda},\ and\ \citenamefont
  {Hashitsume}}]{kubo2012statistical}%
  \BibitemOpen
  \bibfield  {author} {\bibinfo {author} {\bibfnamefont {R.}~\bibnamefont
  {Kubo}}, \bibinfo {author} {\bibfnamefont {M.}~\bibnamefont {Toda}}, \ and\
  \bibinfo {author} {\bibfnamefont {N.}~\bibnamefont {Hashitsume}},\
  }\href@noop {} {\emph {\bibinfo {title} {Statistical physics II:
  nonequilibrium statistical mechanics}}},\ Vol.~\bibinfo {volume} {31}\
  (\bibinfo  {publisher} {Springer Science \& Business Media},\ \bibinfo {year}
  {2012})\BibitemShut {NoStop}%
\bibitem [{\citenamefont {Casati}\ and\ \citenamefont
  {Prosen}(2003)}]{casati2003anomalous}%
  \BibitemOpen
  \bibfield  {author} {\bibinfo {author} {\bibfnamefont {G.}~\bibnamefont
  {Casati}}\ and\ \bibinfo {author} {\bibfnamefont {T.}~\bibnamefont
  {Prosen}},\ }\href@noop {} {\bibfield  {journal} {\bibinfo  {journal}
  {Physical Review E}\ }\textbf {\bibinfo {volume} {67}},\ \bibinfo {pages}
  {015203(R)} (\bibinfo {year} {2003})}\BibitemShut {NoStop}%
\bibitem [{\citenamefont {Das}\ and\ \citenamefont {Dhar}(2014)}]{das2014role}%
  \BibitemOpen
  \bibfield  {author} {\bibinfo {author} {\bibfnamefont {S.~G.}\ \bibnamefont
  {Das}}\ and\ \bibinfo {author} {\bibfnamefont {A.}~\bibnamefont {Dhar}},\
  }\href@noop {} {\bibfield  {journal} {\bibinfo  {journal} {arXiv preprint
  arXiv:1411.5247}\ } (\bibinfo {year} {2014})}\BibitemShut {NoStop}%
\bibitem [{\citenamefont {Gendelman}\ and\ \citenamefont
  {Savin}(2014)}]{gendelman2014normal}%
  \BibitemOpen
  \bibfield  {author} {\bibinfo {author} {\bibfnamefont {O.~V.}\ \bibnamefont
  {Gendelman}}\ and\ \bibinfo {author} {\bibfnamefont {A.~V.}\ \bibnamefont
  {Savin}},\ }\href@noop {} {\bibfield  {journal} {\bibinfo  {journal} {EPL
  (Europhysics Letters)}\ }\textbf {\bibinfo {volume} {106}},\ \bibinfo {pages}
  {34004} (\bibinfo {year} {2014})}\BibitemShut {NoStop}%
\bibitem [{\citenamefont {Mai}\ and\ \citenamefont
  {Narayan}(2006)}]{mai2006universality}%
  \BibitemOpen
  \bibfield  {author} {\bibinfo {author} {\bibfnamefont {T.}~\bibnamefont
  {Mai}}\ and\ \bibinfo {author} {\bibfnamefont {O.}~\bibnamefont {Narayan}},\
  }\href@noop {} {\bibfield  {journal} {\bibinfo  {journal} {Physical Review
  E}\ }\textbf {\bibinfo {volume} {73}},\ \bibinfo {pages} {061202} (\bibinfo
  {year} {2006})}\BibitemShut {NoStop}%
\bibitem [{\citenamefont {G.~R.}\ and\ \citenamefont
  {Barik}(2019)}]{barik2019temperature}%
  \BibitemOpen
  \bibfield  {author} {\bibinfo {author} {\bibfnamefont {A.}~\bibnamefont
  {G.~R.}}\ and\ \bibinfo {author} {\bibfnamefont {D.}~\bibnamefont {Barik}},\
  }\href {\doibase 10.1103/PhysRevE.99.022103} {\bibfield  {journal} {\bibinfo
  {journal} {Phys. Rev. E}\ }\textbf {\bibinfo {volume} {99}},\ \bibinfo
  {pages} {022103} (\bibinfo {year} {2019})}\BibitemShut {NoStop}%
\bibitem [{\citenamefont {Landau}\ and\ \citenamefont
  {Lifshitz}(1987)}]{landau1987fluid}%
  \BibitemOpen
  \bibfield  {author} {\bibinfo {author} {\bibfnamefont {L.}~\bibnamefont
  {Landau}}\ and\ \bibinfo {author} {\bibfnamefont {E.}~\bibnamefont
  {Lifshitz}},\ }\href@noop {} {\enquote {\bibinfo {title} {Fluid mechanics (;
  oxford)},}\ } (\bibinfo {year} {1987})\BibitemShut {NoStop}%
\bibitem [{\citenamefont {Hoover}(2012)}]{hoover2012computational}%
  \BibitemOpen
  \bibfield  {author} {\bibinfo {author} {\bibfnamefont {W.~G.}\ \bibnamefont
  {Hoover}},\ }\href@noop {} {\emph {\bibinfo {title} {Computational
  statistical mechanics}}}\ (\bibinfo  {publisher} {Elsevier},\ \bibinfo {year}
  {2012})\BibitemShut {NoStop}%
\bibitem [{\citenamefont {Saito}\ and\ \citenamefont
  {Dhar}(2011)}]{saito2011additivity}%
  \BibitemOpen
  \bibfield  {author} {\bibinfo {author} {\bibfnamefont {K.}~\bibnamefont
  {Saito}}\ and\ \bibinfo {author} {\bibfnamefont {A.}~\bibnamefont {Dhar}},\
  }\href@noop {} {\bibfield  {journal} {\bibinfo  {journal} {Physical review
  letters}\ }\textbf {\bibinfo {volume} {107}},\ \bibinfo {pages} {250601}
  (\bibinfo {year} {2011})}\BibitemShut {NoStop}%
\bibitem [{\citenamefont {Bedeaux}\ and\ \citenamefont
  {Mazur}(1974)}]{bedeaux1974renormalization}%
  \BibitemOpen
  \bibfield  {author} {\bibinfo {author} {\bibfnamefont {D.}~\bibnamefont
  {Bedeaux}}\ and\ \bibinfo {author} {\bibfnamefont {P.}~\bibnamefont
  {Mazur}},\ }\href@noop {} {\bibfield  {journal} {\bibinfo  {journal}
  {Physica}\ }\textbf {\bibinfo {volume} {73}},\ \bibinfo {pages} {431}
  (\bibinfo {year} {1974})}\BibitemShut {NoStop}%
\bibitem [{\citenamefont {Eyink}(1994)}]{eyink1994renormalization}%
  \BibitemOpen
  \bibfield  {author} {\bibinfo {author} {\bibfnamefont {G.~L.}\ \bibnamefont
  {Eyink}},\ }\href@noop {} {\bibfield  {journal} {\bibinfo  {journal} {Physics
  of Fluids}\ }\textbf {\bibinfo {volume} {6}},\ \bibinfo {pages} {3063}
  (\bibinfo {year} {1994})}\BibitemShut {NoStop}%
\bibitem [{\citenamefont {Drossel}\ and\ \citenamefont
  {Kardar}(2002)}]{drossel2002passive}%
  \BibitemOpen
  \bibfield  {author} {\bibinfo {author} {\bibfnamefont {B.}~\bibnamefont
  {Drossel}}\ and\ \bibinfo {author} {\bibfnamefont {M.}~\bibnamefont
  {Kardar}},\ }\href@noop {} {\bibfield  {journal} {\bibinfo  {journal}
  {Physical Review B}\ }\textbf {\bibinfo {volume} {66}},\ \bibinfo {pages}
  {195414} (\bibinfo {year} {2002})}\BibitemShut {NoStop}%
\bibitem [{\citenamefont {Kardar}\ \emph {et~al.}(1986)\citenamefont {Kardar},
  \citenamefont {Parisi},\ and\ \citenamefont {Zhang}}]{kardar1986dynamic}%
  \BibitemOpen
  \bibfield  {author} {\bibinfo {author} {\bibfnamefont {M.}~\bibnamefont
  {Kardar}}, \bibinfo {author} {\bibfnamefont {G.}~\bibnamefont {Parisi}}, \
  and\ \bibinfo {author} {\bibfnamefont {Y.-C.}\ \bibnamefont {Zhang}},\
  }\href@noop {} {\bibfield  {journal} {\bibinfo  {journal} {Physical Review
  Letters}\ }\textbf {\bibinfo {volume} {56}},\ \bibinfo {pages} {889}
  (\bibinfo {year} {1986})}\BibitemShut {NoStop}%
\bibitem [{\citenamefont {Xiong}\ \emph {et~al.}(2017)\citenamefont {Xiong},
  \citenamefont {Saadatmand},\ and\ \citenamefont
  {Dmitriev}}]{xiong2017crossover}%
  \BibitemOpen
  \bibfield  {author} {\bibinfo {author} {\bibfnamefont {D.}~\bibnamefont
  {Xiong}}, \bibinfo {author} {\bibfnamefont {D.}~\bibnamefont {Saadatmand}}, \
  and\ \bibinfo {author} {\bibfnamefont {S.~V.}\ \bibnamefont {Dmitriev}},\
  }\href@noop {} {\bibfield  {journal} {\bibinfo  {journal} {Physical Review
  E}\ }\textbf {\bibinfo {volume} {96}},\ \bibinfo {pages} {042109} (\bibinfo
  {year} {2017})}\BibitemShut {NoStop}%
\bibitem [{\citenamefont {Rieder}\ \emph {et~al.}(1967)\citenamefont {Rieder},
  \citenamefont {Lebowitz},\ and\ \citenamefont {Lieb}}]{rieder1967properties}%
  \BibitemOpen
  \bibfield  {author} {\bibinfo {author} {\bibfnamefont {Z.}~\bibnamefont
  {Rieder}}, \bibinfo {author} {\bibfnamefont {J.~L.}\ \bibnamefont
  {Lebowitz}}, \ and\ \bibinfo {author} {\bibfnamefont {E.}~\bibnamefont
  {Lieb}},\ }\href@noop {} {\bibfield  {journal} {\bibinfo  {journal} {Journal
  of Mathematical Physics}\ }\textbf {\bibinfo {volume} {8}},\ \bibinfo {pages}
  {1073} (\bibinfo {year} {1967})}\BibitemShut {NoStop}%
\bibitem [{\citenamefont {Spohn}(2014)}]{spohn2014nonlinear}%
  \BibitemOpen
  \bibfield  {author} {\bibinfo {author} {\bibfnamefont {H.}~\bibnamefont
  {Spohn}},\ }\href@noop {} {\bibfield  {journal} {\bibinfo  {journal} {Journal
  of Statistical Physics}\ }\textbf {\bibinfo {volume} {154}},\ \bibinfo
  {pages} {1191} (\bibinfo {year} {2014})}\BibitemShut {NoStop}%
\bibitem [{\citenamefont {Aitchison}\ and\ \citenamefont
  {Hey}(2012)}]{aitchison2012gauge}%
  \BibitemOpen
  \bibfield  {author} {\bibinfo {author} {\bibfnamefont {I.~J.}\ \bibnamefont
  {Aitchison}}\ and\ \bibinfo {author} {\bibfnamefont {A.~J.}\ \bibnamefont
  {Hey}},\ }\href@noop {} {\emph {\bibinfo {title} {Gauge Theories in Particle
  Physics: A Practical Introduction, Volume 2: Non-Abelian Gauge Theories: QCD
  and The Electroweak Theory}}},\ Vol.~\bibinfo {volume} {2}\ (\bibinfo
  {publisher} {CRC Press},\ \bibinfo {year} {2012})\BibitemShut {NoStop}%
\bibitem [{\citenamefont {Di~Cintio}\ \emph {et~al.}(2019)\citenamefont
  {Di~Cintio}, \citenamefont {Iubini}, \citenamefont {Lepri},\ and\
  \citenamefont {Livi}}]{di2019equilibrium}%
  \BibitemOpen
  \bibfield  {author} {\bibinfo {author} {\bibfnamefont {P.}~\bibnamefont
  {Di~Cintio}}, \bibinfo {author} {\bibfnamefont {S.}~\bibnamefont {Iubini}},
  \bibinfo {author} {\bibfnamefont {S.}~\bibnamefont {Lepri}}, \ and\ \bibinfo
  {author} {\bibfnamefont {R.}~\bibnamefont {Livi}},\ }\href@noop {} {\bibfield
   {journal} {\bibinfo  {journal} {Journal of Physics A: Mathematical and
  Theoretical}\ }\textbf {\bibinfo {volume} {52}},\ \bibinfo {pages} {274001}
  (\bibinfo {year} {2019})}\BibitemShut {NoStop}%
\bibitem [{\citenamefont {Aoki}\ and\ \citenamefont
  {Kusnezov}(2001)}]{aoki2001fermi}%
  \BibitemOpen
  \bibfield  {author} {\bibinfo {author} {\bibfnamefont {K.}~\bibnamefont
  {Aoki}}\ and\ \bibinfo {author} {\bibfnamefont {D.}~\bibnamefont
  {Kusnezov}},\ }\href@noop {} {\bibfield  {journal} {\bibinfo  {journal}
  {Physical review letters}\ }\textbf {\bibinfo {volume} {86}},\ \bibinfo
  {pages} {4029} (\bibinfo {year} {2001})}\BibitemShut {NoStop}%
\bibitem [{\citenamefont {Savin}\ \emph {et~al.}(2016)\citenamefont {Savin},
  \citenamefont {Zolotarevskiy},\ and\ \citenamefont
  {Gendelman}}]{savin2016normal}%
  \BibitemOpen
  \bibfield  {author} {\bibinfo {author} {\bibfnamefont {A.~V.}\ \bibnamefont
  {Savin}}, \bibinfo {author} {\bibfnamefont {V.}~\bibnamefont
  {Zolotarevskiy}}, \ and\ \bibinfo {author} {\bibfnamefont {O.~V.}\
  \bibnamefont {Gendelman}},\ }\href@noop {} {\bibfield  {journal} {\bibinfo
  {journal} {EPL (Europhysics Letters)}\ }\textbf {\bibinfo {volume} {113}},\
  \bibinfo {pages} {24003} (\bibinfo {year} {2016})}\BibitemShut {NoStop}%
\bibitem [{\citenamefont {Matsuda}\ and\ \citenamefont
  {Ishii}(1970)}]{matsuda1970localization}%
  \BibitemOpen
  \bibfield  {author} {\bibinfo {author} {\bibfnamefont {H.}~\bibnamefont
  {Matsuda}}\ and\ \bibinfo {author} {\bibfnamefont {K.}~\bibnamefont
  {Ishii}},\ }\href@noop {} {\bibfield  {journal} {\bibinfo  {journal}
  {Progress of Theoretical Physics Supplement}\ }\textbf {\bibinfo {volume}
  {45}},\ \bibinfo {pages} {56} (\bibinfo {year} {1970})}\BibitemShut {NoStop}%
\bibitem [{\citenamefont {Dhar}(2001)}]{dhar2001heat}%
  \BibitemOpen
  \bibfield  {author} {\bibinfo {author} {\bibfnamefont {A.}~\bibnamefont
  {Dhar}},\ }\href@noop {} {\bibfield  {journal} {\bibinfo  {journal} {Physical
  review letters}\ }\textbf {\bibinfo {volume} {86}},\ \bibinfo {pages} {5882}
  (\bibinfo {year} {2001})}\BibitemShut {NoStop}%
\end{thebibliography}
%

\end{document}